\documentclass[10pt,twocolumn,amsmath,amssymb,floatfix,notitlepage,superscriptaddress,longbibliography]{revtex4-1}
\usepackage[T1]{fontenc}
\usepackage[utf8]{inputenc}
\usepackage{lmodern}
\usepackage{microtype,bm,bbm,mathtools}
\usepackage{graphicx,booktabs}
\usepackage{times}
\usepackage[usenames,dvipsnames]{xcolor}

\graphicspath{{figs/}}

\usepackage{hyperref}
\hypersetup{
  colorlinks,
  allcolors=NavyBlue,
  linktoc=all
}

\usepackage[capitalize]{cleveref}
\crefformat{section}{Sec.~#2#1#3}

\begin{document}

\newcommand*\printtitle{Seasonal epidemic spreading on small-world networks:\\
  Biennial outbreaks and classical discrete time crystals}
\title{\printtitle}

\author{Daniel Malz}
\affiliation{Max-Planck-Institute of Quantum Optics, Hans-Kopfermann-Str. 1, 85748 Garching, Germany}
\affiliation{Munich Center for Quantum Science and Technology, Schellingstra{\ss}e 4, 80799 Munich, Germany}
\author{Andrea Pizzi}
\affiliation{Cavendish Laboratory, University of Cambridge, Cambridge CB3 0HE, United Kingdom}
\author{Andreas Nunnenkamp}
\affiliation{School of Physics and Astronomy and Centre for the Mathematics and Theoretical Physics of Quantum Non-Equilibrium Systems, University of Nottingham, Nottingham, NG7 2RD, United Kingdom}
\author{Johannes Knolle}
\affiliation{Department of Physics, Technische Universit{\"a}t M{\"u}nchen, James-Franck-Stra{\ss}e 1, D-85748 Garching, Germany}
\affiliation{Munich Center for Quantum Science and Technology, Schellingstra{\ss}e 4, 80799 Munich, Germany}
\affiliation{Blackett Laboratory, Imperial College London, London SW7 2AZ, United Kingdom}

\begin{abstract}
  We study seasonal epidemic spreading in a susceptible-infected-removed-susceptible (SIRS) model on small-world graphs. We derive a mean-field description that accurately captures the salient features of the model, most notably a phase transition between annual and biennial outbreaks. A numerical scaling analysis exhibits a diverging autocorrelation time in the thermodynamic limit, which confirms the presence of a classical discrete time crystalline phase. We derive the phase diagram of the model both from mean-field theory and from numerics. Our work offers new perspectives by demonstrating that small-worldness and non-Markovianity can stabilize a classical discrete time crystal, 
  and by linking recent efforts to understand such dynamical phases of matter to
  the century-old problem of biennial epidemics.
\end{abstract}
\maketitle

\section{Introduction}
Already its inventor, Daniel Bernoulli, recognized the use of epidemic modelling to guide public health decisions by advocating inoculations to prevent the spread of smallpox~\cite{Bernoulli1760}.
Naturally, understanding and preventing disease has always been of great interest, reflected in the correspondingly vast body of literature~\cite{Anderson1992,KeelingAndRohani2011}.
Pioneered in the early $20^{\mathrm{th}}$ century~\cite{Hamer1906,Kermack1927,Soper1929}, a modern formulation of epidemic dynamics uses coupled ordinary differential equations for the number of susceptible (S), exposed (E), infected (I), and recovered (R) individuals.
Owing to their importance, SIR models have been an early target for computer-aided simulations~\cite{London1973}, and to this day, sophisticated versions of SEIRS models, accounting for factors such as seasonality, immunity, cross-immunity between virus strains, and the effect of distancing measures, are used to model epidemics~\cite{Kissler2020}.

\begin{figure}[t]
  \centering
  \includegraphics[width=\linewidth]{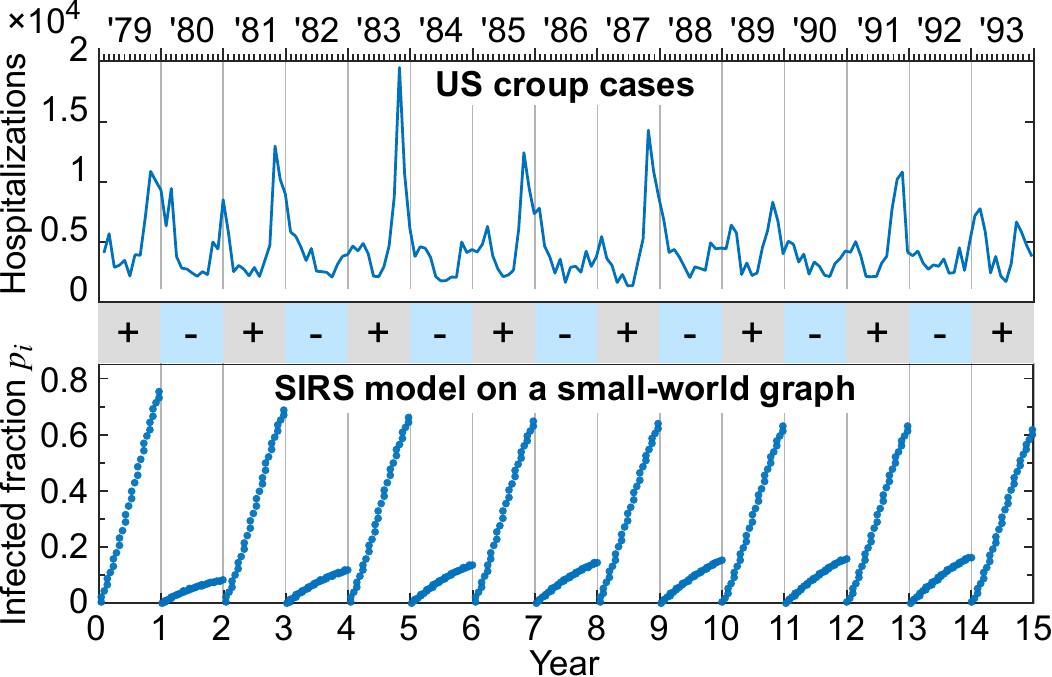}
  \caption{
	\textbf{Biennial epidemics.} 
	Biennial outbreaks appear in actual data of croup cases in the US over several decades, reflected in the number of hospitalizations~\cite{[{Croup case data were obtained from the National Hospital Discharge Survey (NHDS) according to the procedures used in }][{, which previously plotted the same data.}]marx1997pediatric}.
	A qualitatively similar behavior appears in our seasonal SIRS model on a small-world graph, in which the fraction of infected vertices $p_i$ also shows biennial major outbreaks (bottom, for $\phi = 5 \times 10^{-3}$, $k = 2$ and $N = 10^6$).
	On an abstract level, this subharmonic response breaks time-translation symmetry, which can be considered a crystalline phase in time (bar with alternating $+-$, corresponding to major and minor outbreaks, respectively).
	Defects in this `crystal' may occur due to noise and perturbations and destroy the crystalline order.
  }
  \label{fig:1}
\end{figure}

Among the many rich dynamical phenomena that can be captured by S(E)I(RS) models is the emergence of biennial outbreaks, as have been observed in real world measles case data already more than a century ago~\cite{Hamer1906}. This behaviour of epidemic spreading is subharmonic, in the sense that the periodicity of the outbreaks (two years) is a multiple of that of the underlying statistical laws (one year, given through seasonality).
%
At the core of biennial epidemics is the fact that outbreaks arise through a combination of (i) seasonally increased infectivity and (ii) a large enough pool of susceptible individuals. Under suitable circumstances, an outbreak can remove enough susceptible individuals (through death or immunity acquisition) to induce herd immunity preventing outbursts in the subsequent year(s). Only once the pool of susceptible individuals has been replenished (through birth~\cite{Hamer1906} or immunity loss) another outbreak can occur, which results in outbreaks spaced by multiple years.
Indeed, such a behavior is not specific to measles, but has been studied in chickenpox and mumps~\cite{London1973}, croup~\cite{Marx1997} (\emph{cf.} \cref{fig:1}), and has been predicted for COVID-19 as well~\cite{Kissler2020}.
In this paper we take biennial outbreaks as a motivation to study whether this subharmonic response corresponds to robust time-crystalline order in the thermodynamic limit (as defined later), or if it is instead washed out by noise intrinsic to the system.

The elegant and simple description via global variables has had tremendous success, but misses potentially important effects due to spatial fluctuations and the probabilistic nature of the transmission.
This is more appropriately captured by probabilistic cellular automata on networks, where vertices (cells) represent individuals and edges potential infection pathways~\cite{VanMieghem2014,Pastor-Satorras2015}.
Among the key challenges is to understand the limits of the validity of global variables and to characterize phenomena that cannot be described otherwise~\cite{Vespignani2012}.
For example, pandemics are both characterized by large spatial fluctuations and have fat tails~\cite{Cirillo2020}, important aspects missed by the global-variables approach, but naturally captured by scale-free networks~\cite{Barabasi1999,Pastor-Satorras2001,May2001,Pastor-Satorras2002}.
Such effects play a large role in immunization strategies~\cite{Goldenberg2005}.

A natural choice is to study infection dynamics on small-world graphs~\cite{Moore2000}, as they exhibit the famous small-world property that characterizes social networks~\cite{Milgram1967}, for which the average distance between any two vertices is several orders of magnitude smaller than the size of the network~\cite{Watts1998}.
SI(R) models on small-world graphs allow for analytic understanding in the large-system-size limit~\cite{Newman2000,Hastings2003} and naturally predict the well-known phenomenon of herd immunity, which is linked to the presence or absence of a percolating cluster~\cite{Moore2000}.

\begin{figure*}[htb]
  \centering
  \includegraphics[width=\linewidth]{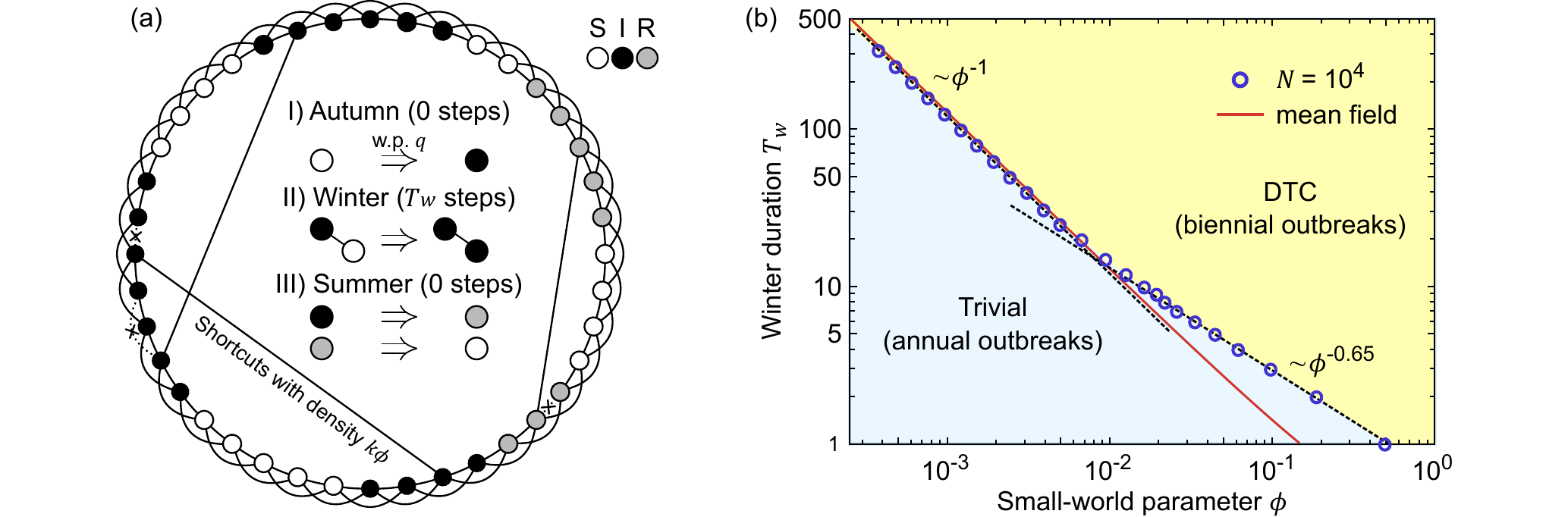}
  \caption{
	\textbf{Seasonal epidemic spreading on small-world graphs.}
	(a) Schematic of the model for $N = 40$ vertices. The graph is constructed starting from a periodic one-dimensional lattice in which each vertex is connected to its $2k$ nearest neighbors. In a `rewiring' procedure, with probability $\phi$, the ends of each short-range edge (dashed) can be moved to a random location in the graph, creating a shortcut. Each vertex is either susceptible (S, white), infected (I, black), or recovered (R, grey), that is, immune. Disease spreading occurs according to the shown three update rules, that are further detailed in the main text and in \cref{app:numerics}.
	(b) Phase diagram in the plane of the small-world parameter $\phi$ and winter duration $T_w$, comparing numerical results for $N = 10^4$ and $k=2$ with the mean-field prediction. The trivial and DTC phases are characterized by annual or biennial outbreaks, respectively.
	We fit $T_w=\alpha\phi^{\eta}$ to the first eight and last eight values (dashed) and find $(\alpha,\eta)=(0.12\pm.01,-1\pm.01)$ and $(.65\pm.02,-.65\pm.01)$.
	The mean-field prediction (solid red) shows striking agreement with numerics for small $\phi$, but cannot explain the deviation at large $\phi$, which we attribute to a number of effects that are missed by mean-field theory, most notably the discreteness of space and time.
  }
  \label{fig:2}
\end{figure*}

From the point of view of statistical mechanics, the genuinely theoretical question whether biennial epidemics are just transient phenomena or can rather persist up to \emph{infinite} times is a very interesting and highly non-trivial one. The subharmonic response observed in structureless mean-field models is generally spoiled when moving to a network, where noise tends to randomize the phase of the response and to ultimately destroy any system's autocorrelation with its far past.
For example in the case of epidemic outbreaks, a decay of the autocorrelation of the system would occur if in two consecutive years either two large outbreaks or no outbreaks occur, as this would correspond to outbreaks shifting from odd to even years. In the interpretation in terms of crystalline order in time (see  \cref{fig:1}), this occurrence would introduce a defect. 
It is natural to expect defects to occur in the presence of noise and perturbations, but recently it was discovered that so-called `classical discrete time crystals' (DTCs) are a dynamical phase of matter that nevertheless maintains a persistent subharmonic response in the thermodynamic limit.
DTCs have originally been studied in the context of quantum many-body localization~\cite{Khemani2016, Else2016, Yao2017, Moessner2017} and have in the last few years attracted a tremendous amount of interest \cite{Sacha2015,Choi2017, Zhang2017, Gong2018a, Pizzi2019, Machado2020}.
Recent work has extended the field to classical stochastic Markovian processes in one~\cite{Yao2020,Pizzi2020} and two dimensions~\cite{Gambetta2019}.

Here, we investigate a seasonal SIRS model on small-world graphs and show under which circumstances it gives rise to biennial outbreaks, supporting numerics with an analytical mean-field theory that captures the salient features of the model's rich phenomenology. 
With a scaling analysis, we show that in the thermodynamic limit, biennial outbreaks exhibit long-range order in time, heralded by a diverging autocorrelation time and a corresponding exponential suppression of defects. We attribute the robustness of the DTC to the interplay of small-worldness and, crucially, non-Markovianity.
Our work links the well-known phenomena of biennial outbreaks to the theory of classical DTCs and thus unveils striking real-world ramifications of these remarkable dynamical phases of matter.

The remainder of this paper is organized as follows. In Section II we introduce the model, detailing the structure of the small-world network and the update rules describing seasonal epidemic spreading, and present its dynamical phase diagram. In Section III we gain understanding of the system by deriving an analytical mean-field theory.
In Section IV we present a thorough characterization of the model and we investigate the time-crystalline nature of the biennial epidemics by means of a scaling analysis to assess its stability. Finally, in Section V we discuss the results and conclude with an outlook for future research.

\section{Model}

We consider a stochastic microscopic SIRS model for seasonal epidemic spreading on the celebrated small-world networks introduced by Watts and Strogatz~\cite{Watts1998}, sketched and described in \cref{fig:2}a. The shortcut density $\phi$ parameterizes the `small-worldness', which allows us to interpolate in a controlled manner from a linear 1d chain ($\phi = 0$) to random graphs ($\phi = 1$). Each vertex can be in one of three states: susceptible (S), infected (I), immune/recovered (R), and the system evolves according to the following rules:
\begin{enumerate}
  \item[(I)] Autumn: vertices are exposed to the disease with a probability $q$ and, if susceptible, contract it, S$\to$I.
  \item[(II)] Winter: in each of total $T_w$ steps, the infection spreads from infected vertices to susceptible vertices they are connected to, S$\to$I.
  \item[(III)] Summer: the infected vertices recover and become immune, I$\to$R, whereas immune vertices lose their immunity and become susceptible again, R$\to$S.
\end{enumerate}
These rules are repeated annually, which sets the fundamental periodicity of the model.
In the first year we take all vertices to be susceptible, but the qualitative results shown here are robust to perturbations of the initial condition.
An operational definition of how we implemented these rules is given in \cref{app:numerics}.

The motivation for this model is that, in a social network, long-distance travellers act as shortcuts for disease spreading across distant regions of the globe, as parameterized by the small-world parameter $\phi$. Similarly, we can imagine that in autumn the vertices are exposed to the infection because of travellers returning from hidden regions (\emph{e.g.}, faraway countries) that are not accounted for explicitly in the network, so that $q \sim \phi$. In the following, we make the choice $q = \phi$ for concreteness.
The ``winter time'' $T_w$ has been introduced here as an artificial way to stop the infection, as there is otherwise no mechanism to stop the propagation of the disease, which would make the dynamics trivially biennial. The motivation for this and imagined underlying mechanism is a presumed seasonally varying infectivity, which eventually causes the outbreak to end.
We have chosen here to take the loss of immunity to happen exactly after one year, which fundamentally is the reason that the model exhibits a biennial response. If immunity would last for $n$ years, we expect the outbreaks to be spaced by $n+1$ years. 
We stress that our goal here is not to model real-world data, but instead to study a minimal toy model to explore the statistical mechanics of epidemic spreading. Nevertheless, the results are expected to extend beyond the basic model studied here.

Depending on the value of the parameters $T_w$ and $\phi$, at long times the system can behave in two fundamentally different ways, featuring outbreaks that occur either annually or biennially. As we will show later, these two behaviors constitute two genuine dynamical phases of matter: The phase with annual outbreaks preserves the time translation symmetry of the model, and can therefore be referred to as the `trivial' phase. In contrast, if outbreaks occur biennially, the discrete time translation symmetry is broken, and the system enters the DTC phase. The phase diagram, derived analytically and confirmed numerically, is shown in \cref{fig:1}d.
Intuitively, the transition may be understood in terms of the effect of immune vertices.
After a long winter ($T_w$ large) and/or a fast outbreak ($\phi$ large), in which a large proportion of vertices has been infected, the established herd immunity suppresses the outbreak in the following year. Two years later, the immunity has decayed, and a large outbreak occurs again.
This intuitive understanding is built in the limit in which the infection reaches the whole connected cluster of susceptible individuals (\emph{e.g.}, infinite $T_w$). In more realistic scenarios, not all susceptible individuals will get infected, which is captured through a finite $T_w$. Nevertheless, as we show below, the subharmonic response is stable down to some critical $T_w$, below which it disappears.
Underlying these effects is clearly the immunity, which prevents a given vertex to be infected for two consecutive years, a memory effect that we refer to as `non-Markovianity'. Since the infection spreading is quicker on a smaller world, the critical $\phi_c$ dividing the two phases decreases with the winter duration $T_w$.
We fit $\alpha \phi^\eta$ to the data (shown as dashed lines in \cref{fig:2}b and obtain the critical exponents $\eta=-1\pm.01$ for small $\phi$, and $\eta=.65\pm.01$ for large $\phi$.
At small $\phi$ the agreement with mean-field theory is striking, whereas for large $\phi$ the discreteness of the model plays an increasing role, which cannot be accounted for by mean field.

\section{Mean-field solution}
To gain analytical understanding of the trivial phase, the DTC, and the transition between them, we derive a mean-field theory. To do so, we first find analytical expressions for the epidemic dynamics throughout a single winter, and then describe how the epidemic evolves from one year to the next.
\subsection{Single-year dynamics}
Pioneered by Newman \emph{et al.}~\cite{Newman2000}, an accurate mean-field description can be found in the limit of large graphs $N \to \infty$ and small small-world parameter $\phi$, by treating space and time as continuous, such that the epidemic dynamics can be described by ordinary differential equations. In contrast to the approach in Ref.~\cite{Newman2000}, here we have to account for the presence of immune vertices that act to suppress outbreaks, as this underpins the biennial response. We do so while retaining the analytic solubility of the model by making the approximation that immune vertices are randomly distributed. Thus, they can be accounted for by the probability $P_{S|\bar{I}}$ that a non-infected vertex is susceptible (rather than immune)
\begin{equation}
  P_{S|\bar{I}}(t) = 1 -\frac{p_r}{1-p_i(t)},
  \label{eq:PSI}
\end{equation}
where $p_i(t)$ and $p_r$ are the fractions of infected and immune vertices, respectively. Since immunity only changes in summer (\emph{cf.} \cref{fig:2}a), $p_r$ is constant throughout the winter.

In the following we mean by `infected region' a contiguous local chain (connected by short-range edges) of infected vertices. In contrast to the concept of connected clusters, we therefore count two regions connected by a long-range edge as two separate regions. For instance, there are three regions in~\cref{fig:2}a. Let us denote the density of infected regions (the fraction of infected vertices) by $\nu_i(t)$. In every time step each of the $2N\nu_i$ infection fronts (boundaries between infected and susceptible regions) advances by $k$ steps, enveloping $2Nk\nu_i$ vertices. To obtain the rate at which sites are infected, we have to multiply this by the probability that the enveloped vertices are susceptible,
\begin{equation}
  \frac{d p_i(t)}{dt} = 2k\nu_i P_{S|\bar{I}}.
  \label{eq:dpi}
\end{equation}

In the spreading of infection fronts, the number of infected regions may change according to two mechanisms. One, a new infected region is spawned if an infection front crosses a shortcut between two susceptible vertices. The rate for this happening is the product of the vertices exposed in one time step, $2Nk\nu_i$, and the density of shortcuts among non-infected regions, $2k\phi(1-p_i)$.
The factor of $1-p_i$ arises, as the remaining shortcuts necessarily connect two uninfected regions.
Two, infected regions can merge, which decreases $\nu_i$ at a rate of $2k\nu_i^2/(1-p_i)$.
This factor is most easily understood as the product of the rate at which the total susceptible regions shrink ($2k\nu_i$) times the density of boundaries to infected regions in the remaining susceptible fraction ($\nu_i/(1-p_i)$).
Optionally, the distribution of gap sizes can be considered explicitly~\cite{Newman2000}.
Crucially, both these processes will be effective only if the involved non-infected vertices are susceptible, which happens with probability $P_{S|\bar{I}}(t)$, such that
\begin{equation}
  \frac{d\nu_i}{dt} = \left( 4k^2 \phi (1-p_i) \nu_i - \frac{2k \nu_i^2}{1-p_i} \right) P_{S|\bar{I}}.
  \label{eq:dnui}
\end{equation}
For $P_{S|\bar I}=1$ (that is, no immune sites) \cref{eq:dpi,eq:dnui} reduce to the equations introduced by Newman, Moore and Watts~\cite{Newman2000}. Here, however, we want to consider the dependence of $P_{S|\bar I}$ on the number of immune ($p_r$) and infected ($p_i$) sites.
Such a dependence generally makes \cref{eq:dpi,eq:dnui} unsolvable.
Remarkably, solvability is recovered for the specific choice in \cref{eq:PSI}, for which we integrate analytically \cref{eq:dpi,eq:dnui} (\emph{cf.} \cref{app:mean-field}), yielding
\begin{equation}
  p_i(t) = (1-p_r)\frac{e^{2t\theta}-(1-\phi)(1-2k\phi)}{e^{2t\theta}+2k(1-\phi)} \equiv f(p_r, t),
  \label{eq:mf}
\end{equation}
where the notation $f(p_r, t)$ emphasizes the dependence on the immune fraction $p_r$, and where the inverse timescale $\theta$ reads
\begin{equation}
  \theta=\phi k(1-p_r)(1+2k(1-\phi)).
  \label{eq:theta}
\end{equation}

As a sanity check, in the limit $t \to 0$ we obtain $p_i = \phi(1-p_r)$, recovering the initial fraction of infected vertices,
whereas in the limit $t \to \infty$ the fraction of infected vertices asymptotically approaches $p_i \to 1-p_r$.

\begin{figure*}[htb]
  \centering
  \includegraphics[width=\linewidth]{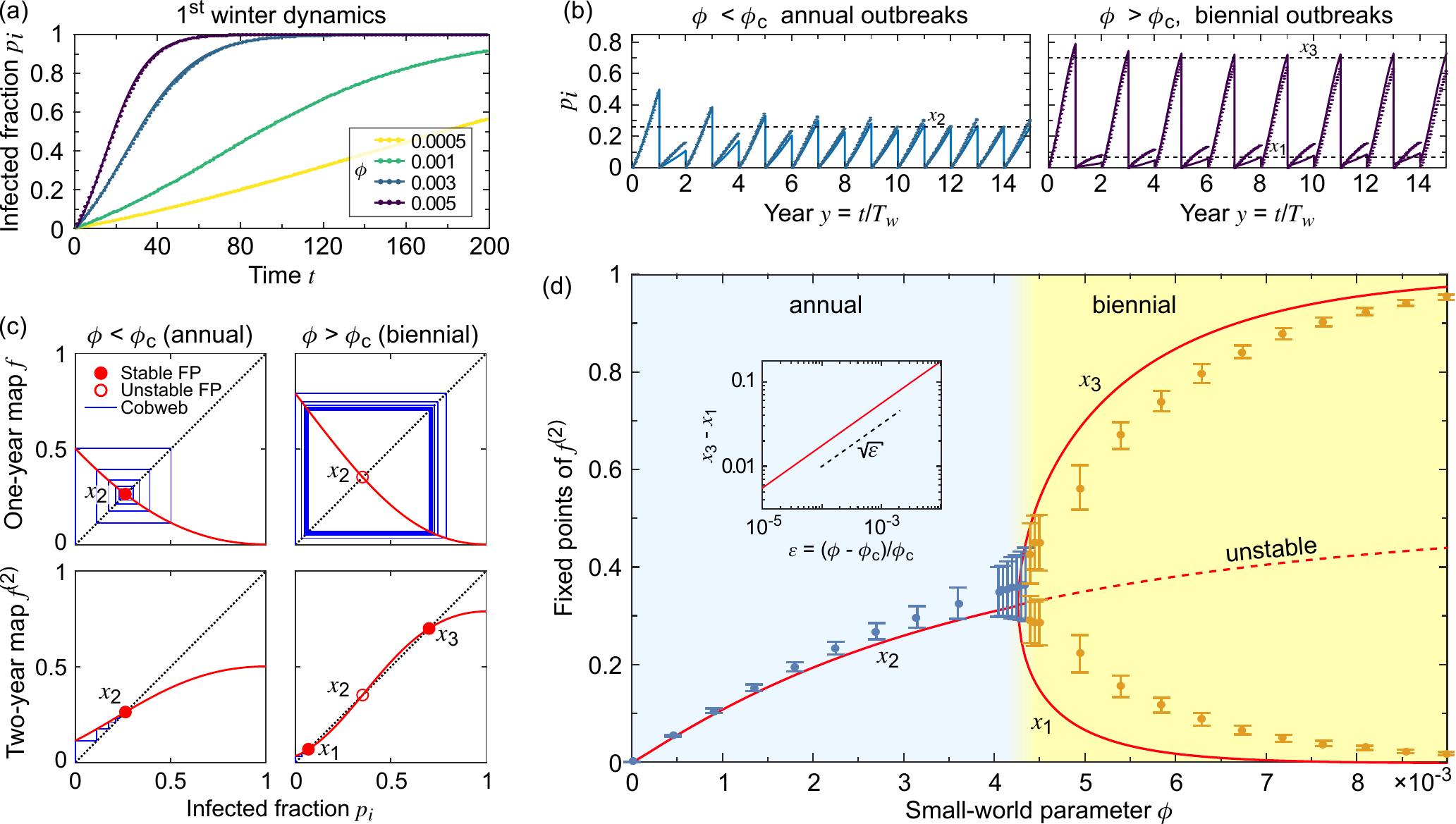}
  \caption{
	\textbf{Characterization of the epidemic dynamics.}
	(a) Time evolution of the infected fraction $p_i$ throughout a single winter. Considering no immune vertices in the network, we observe a logistic-like growth, with an initial superlinear rise (quickest for smaller worlds, that is, larger $\phi$) and eventual saturation when all vertices get infected, as it is also accurately reproduced by mean-field.
	(b) Epidemics over multiple years can have two qualitatively very different behaviors. For $\phi < \phi_c$ (left) and $\phi > \phi_c$ (right), outbreaks occur annually and biennially, respectively. In the latter case, corresponding to a small-enough-world, the system responds with a period twice that of the underlying seasonality, pointing to a DTC. Mean-field (solid line) is compared to numerics for $N = 10^4$ (markers).
	(c) Mean-field fixed-points analysis. The one-year map $f$ only has one FP $x_2$. If $\phi < \phi_c$, $x_2$ is stable and the end-of-winter infected fraction $p_i^{(n)}$ eventually reaches $x_2$. On the contrary, if $\phi > \phi_c$, $x_2$ becomes unstable and $p_i^{(n)}$ eventually oscillates at every year between $x_1$ and $x_3$, the emerging stable FPs of the two-year map $f^{(2)}$. The mean-field dynamics is highlighted with a cobweb in blue.
	(d) Bifurcation diagram of the FPs of $f^{(2)}$ within mean-field (red lines) and end-of-winter infected fraction $p_i^{(n)}$ in the even and odd years at long times for $N = 10^5$ (blue markers). Solid and dashed lines stand for stable and unstable FPs, respectively. In mean-field, the DTC order parameter $x_3-x_2$ scales with critical exponent $1/2$ (inset).
	Note that while the mean-field predictions are otherwise very accurate, the critical exponent is likely larger, as can be seen from the discrepancy on the right side of the phase transition in (d). We attribute this primarily to its inability to account for the clusterized structure of the immune vertices. We were not able to get sufficiently good data to make quantitative predictions of the critical exponent though. 
	Mean-field is surprisingly successful in locating the critical $\phi_c$.
  	For (b) and (d) we have chosen $T_w=30$.
  }
  \label{fig:3}
\end{figure*}

\subsection{Multi-year dynamics}
The solution to epidemic spreading in a single year provides us with a formula for the total fraction of infected vertices at the end of one year $ p_i^{(n)} = f(p_r^{(n)}, T_w)$,
where $p_r^{(n)}$ is the immune fraction during the $n$-th winter. The summer transition from infected to immune means that $p_r^{(n)} = p_i^{(n-1)}$, and therefore we obtain the
discrete `one-year map'
\begin{equation}
  p_i^{(n+1)} = f(p_i^{(n)}),
  \label{eq:f}
\end{equation}
where we have omitted the dependence of $f$ on the winter duration $T_w$, which is henceforth treated as a fixed parameter. Applying $f$ twice, we obtain the two-year map $f^{(2)}$
\begin{equation}
  p_i^{(n+2)} = f^{(2)}(p_i^{(n)}) = f(f(p_i^{(n)})),
  \label{eq:f2}
\end{equation}
which is helpful in describing the biennial epidemics.

Note that to leading order in $\phi$, only the product of $T_w$ and $\phi$ appears in $f$,
which readily explains why at small $\phi$, the critical small-world parameter scales as $\phi_c \sim 1/T_w$ (\emph{cf.} \cref{fig:2}(b)).

To conclude this section, we briefly discuss the shortcomings of our mean-field theory.
Most importantly, while the variable $\nu_i$ contains the information on the number of infected regions, no such information is stored for the immune population, which only enters in the form of the probability $P_{S|\bar{I}}$.
This discounts the effect that clusters of immune sites may stop infection fronts, and also enclose a susceptible cluster, preventing it from being infected.
Indeed, the mean-field model does not account for any fluctuations or other random structure that may emerge, such as fluctuations in the number of initially infected vertices, the relative positioning that may or may not be conducive to fast outbreaks, the number of long-range links, the size of the connected cluster of the randomly sampled small-world graph, etc.
Ultimately, the mean-field theory is validated by the good agreement with numerics, which is shown in the next Section.

\section{Results and analysis}
\subsection{Phase diagram}
Throughout a single winter, the fraction of infected vertices $p_i(t)$ follows a characteristic logistic-like growth, with an initial superlinear growth followed by saturation to an asymptotic value at long times. In Fig.~\ref{fig:3}(a) we verify that the mean-field agrees remarkably well with the full numerics in the absence of immune vertices ($p_r = 0$). The multi-year dynamics is investigated in Fig.~\ref{fig:3}(b), which, depending on the small-world parameter $\phi$, exhibits two qualitatively different behaviors. If $\phi$ is smaller than a critical $\phi_c$, we observe annual outbreaks, whereas if $\phi > \phi_c$, major outbreaks occur biennially, and $p_i(t)$ oscillates with a period of two years.

We observe that the accuracy of the mean-field predictions deteriorates after the first winter. To understand why this is the case, we note that a vertex that gets randomly infected at the beginning of the winter typically lies in the bulk of a susceptible region. In the early stages of the winter the disease thus propagates essentially unperturbed by the presence of the immune vertices. Only at a later stage of the winter some infection fronts get stopped by immune regions. This effect depends on the spatial structure and distribution of immune vertices and therefore cannot be captured by mean field. Nonetheless, as we are about to see, mean-field theory is still remarkably accurate in capturing and explaining the phase transition between annual and biennial epidemics.

Within mean-field, in fact, the long-time dynamics can be understood using the tools of discrete dynamical maps~\cite{Gros2013}. A simple inspection as shown in \cref{fig:3}(c) reveals that the equation $x = f(x)$ is always fulfilled by one and just one fixed point (FP), that we call $x_2$. The stability of this FP is determined by the first derivative $f'(x_2)$.
Since the infected fraction decreases with increasing immune fraction, $f'(x) < 0$, and the only possibilities are
\begin{equation}
  \begin{cases}
	f'(x_2) > -1 \quad \Rightarrow \quad x_2 \ \text{is stable} \\
	f'(x_2) < -1 \quad \Rightarrow \quad x_2 \ \text{is unstable}.
  \end{cases}
\end{equation}
When $x_2$ is stable, $p^{(n)} \xrightarrow{n \to \infty} x_2$, and the system lies in the trivial phase with annual outbreaks at long times. To understand what happens when $x_2$ is unstable, we turn to the two-year map $f^{(2)}$ in Eq.~\eqref{eq:f2}, which is also investigated in Fig.~\ref{fig:3}(c). As it is easy to check, $x_2$ is a FP also of $f^{(2)}$. When the FP $x_2$ is unstable, $f^{(2)\prime}(x_2)=[f'(x_2)]^2 > 1$, indicating the emergence of two new stable FPs $x_1$ and $x_3$ for $f^{(2)}$.
When $x_2$ is unstable, at long times we therefore get $p_i^{(2n)} \xrightarrow{n \to \infty} x_1$ and $p_i^{(2n+1)} \xrightarrow{n \to \infty} x_3$ (or vice versa, depending on the initial condition). That is, small and large outbreaks alternate (quantified by $x_1$ and $x_3$, respectively).
The phase boundary occurs at the critical line defined by $f'(x_2) = -1$, with $x_2 = f(x_2)$.

Plotting the FPs of the two-year map $f^{(2)}$ versus the small-world parameter $\phi$, in Fig.~\ref{fig:3}(d) we obtain a bifurcation diagram. At $T_w=30$, the critical small-world parameter that separates the trivial and the DTC phases is $\phi_c = 4.24(8) \times 10^{-3}$. Near criticality, the mean-field scaling $x_3 - x_1 \sim (\phi - \phi_c)^\xi$ is characterized by the critical exponent $\xi=1/2$ (see inset). Numerics suggests that the actual critical exponent may be larger, although large fluctuations close to the transition, a finite-size effect, prevent us from drawing clear conclusions.

Varying the winter duration $T_w$, this method allows us to calculate the full phase diagram shown in \cref{fig:2}b.
The mean-field model predicts the phase separation to lie at $\phi_c=0.126918\, T_w^{-1}$ (dashed line), which shows striking agreement with numerics for small $\phi$, which predicts $\phi_c=0.12065T_w^{-1}$.
As in previous work~\cite{Newman2000}, the approximations made to derive the mean-field solution deteriorate at larger values of $\phi$, which explains the deviation.

\begin{figure}[tb]
	\centering
	\includegraphics[width=\linewidth]{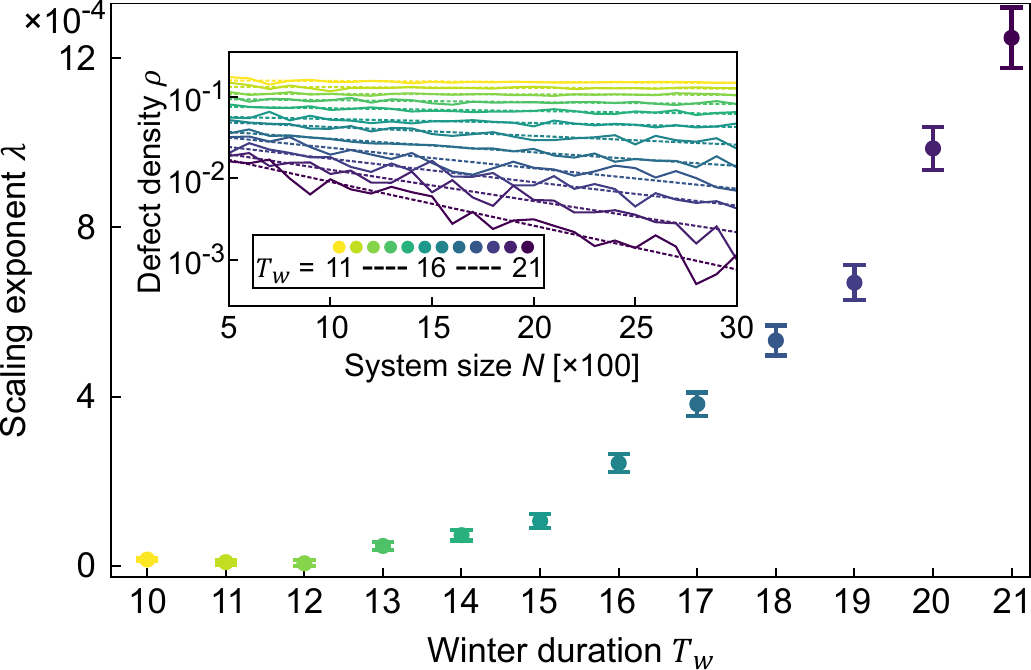}
	\caption{\textbf{Defect density in the time crystal.}
		Inset: The defect density $\rho\propto\exp[-\lambda(T_w) N]$ scales exponentially as a function of system size $N$.
		Main panel: The departure of the scaling exponent $\lambda(T_w)$ signals the transition from trivial to DTC phase.
		Calculated using $\phi=0.01$, $k=2$, and 100 averages of time traces ranging over 10,000 years.
		Note that here we consider much smaller system sizes than in the other figures in order to resolve the scaling of autocorrelation time with reasonable computational effort.
	}
	\label{fig:4}
\end{figure}

\subsection{Phase stability}
It is impossible to determine whether the biennial response is a genuine dynamical \emph{phase} of matter from mean-field theory alone.
The reason is that microscopic fluctuations typically destroy correlations over time, due to the presence of random phase flips (defects in the crystal).
A defect in this setting means that two consecutive years have minor (or major) outbreaks, such that the biennial response is broken.

To signal a DTC, the subharmonic response of the system needs to show a diverging autocorrelation time with system size, and this divergence should be robust both to noise and to the perturbation of the model's parameters~\cite{gambetta2019classical, Yao2020}. To analyze this, we extract the defect density from calculated time traces and observe their scaling with system size (\emph{cf.} \cref{fig:4} inset).
In the trivial phase, we find that defects occur at random irrespective of system size.
This is expected, as there is no subharmonic response in the first place. With increasing winter duration $T_w$, the system crosses into the DTC phase and we observe clear exponential scaling with system size (see inset in \cref{fig:4}). The exponent $\lambda(T_w)$ characterizing this scaling shows a clear departure from zero as the system enters the DTC phase, e.g., at $T_w \approx 13$ in \cref{fig:4}.

\section{Discussion and conclusions}
Revisiting the century-old problem of biennial epidemics from the perspective of statistical mechanics, we have studied a microscopic model of seasonal epidemic spreading on a small-world network, giving rise under appropriate circumstances to a biennial outbreak behavior, which is well-described by a fully analytical mean-field theory. Surprisingly, the subharmonic response of the stochastic dynamical system is stable in the thermodynamic limit against noise and perturbations, and we therefore classify it as a classical DTC.

Previous research highlighted that classical DTCs are typically prevented by deconfined domain walls in short-range 1d models~\cite{Yao2020}, but can be stabilized by either larger dimensionality~\cite{Gambetta2019} or long-range interactions~\cite{Pizzi2020}.
These works also show how the essential features of classical DTCs can be captured by probabilistic cellular automata (PCA)~\cite{Yao2020,Pizzi2020}, in which the system is described by two-state variables.
This raises the more general question of which aspects (dimensionality, range of interactions, contractive dynamics,..) of these models are responsible for stabilizing the time crystalline phase.
In this context, on the one hand our work explores small-world graphs, in which a few random links shortcut across a 1d chain, accounting for the small-world property of real-world social networks.
On the other hand, a crucial role is played by immunity, which could equivalently be understood as a non-Markovian memory effect that precludes a vertex from being infected two years in a row.
Our work therefore identifies non-Markovianity and small-worldness as two key ingredients that can stabilize time-crystallinity against the proliferation of noise-induced defects.

We conclude outlining some possible directions for future research.
First, it would be desirable to improve the mean-field theory further such that it accounts explicitly for the number $\nu_r$ of immune regions.
Such a theory could start from differential equations analogous to Eqs.~\eqref{eq:dpi} and \eqref{eq:dnui}, but introducing the parameter $\nu_r$ would most likely come at the price of losing analytical solubility.
Second, the model studied here is very simplistic compared to models used to forecast actual epidemics. Additional ingredients may reveal a richer phase diagram or novel phenomena.
As a few examples, one could for instance consider the survival of the infection through the summer (rather than a random autumn exposure), account for different epidemic spreading rules in various regions of the network (e.g., as to mimic different seasonalities in the two hemispheres), include more states of the vertices (e.g., infected asymptomatic) or strains of virus, or account for randomized and/or longer immunity times, which could result in the emergence of $n$-DTCs with major outbreaks every $n$ years. Interestingly, the last point relates to early work that argued that only $n=2$ could produce a stable subharmonic response~\cite{Bennett1990}, in contrast to what our work would suggest.
A particularly timely avenue is to investigate the effects of periodic interventions (such as an intermittent reduction of $\phi$ as to mimic travel restrictions), and whether these can lead to a subharmonic response, too.
Third, more complex, multi-state variable models support even richer dynamics, such as the prime-numbered response in predator-prey models on 2d graphs~\cite{Goles2001}. 
While in our model the subharmonicity is set by the duration of immunity, this raises the possibility of time crystals in which the periodicity is generated dynamically.

Finally, an open question that persists is whether models of the kind studied here may support \emph{continuous} time symmetry breaking. Indeed, in the absence of periodic driving, this model is known to produce a periodic response~\cite{Kuperman2001}, and future investigations should clarify whether the system's autocorrelation time diverges with the system size. The possibility of continuous time symmetry breaking, proven impossible for ground states of time-independent quantum Hamiltonian systems~\cite{Watanabe2015}, opens a new avenue of research in a classical stochastic setting.

\acknowledgements
We acknowledge support from the Imperial-TUM flagship partnership. D.~M.\ acknowledges funding from ERC Advanced Grant QENOCOBA under the EU Horizon 2020 program (Grant Agreement No. 742102).
A.~P.~acknowledges support from the Royal Society. A.~N.~holds a University Research Fellowship from the Royal Society.

\appendix

%
%
%
%
%
\section{Mean-field theory:\\ Solution of the single-winter dynamics}\label{app:mean-field}
Here, we present a step by step solution of the Eqs.~\eqref{eq:dpi} and \eqref{eq:dnui} in the main text. First, we prefer to rewrite the equations for $p_i$ into an equation for the complimentary $\mu_i = 1-p_i$ and $\mu_r = 1-p_r$.
This yields the two coupled equations
\begin{equation}
  \frac{d\mu_i}{dt} = -2 k \nu_i \left(1 -\frac{1-\mu_r}{\mu_i}\right)
  \label{eq dmui SM}
\end{equation}
and
\begin{equation}
  \frac{d\nu_i}{dt} = \left( 4k^2 \phi \mu_i \nu_i - \frac{2 k \nu_i^2}{\mu_i} \right) \left(1 -\frac{1-\mu_r}{\mu_i}\right).
  \label{eq:dnui SM}
\end{equation}
To solve Eqs.~\eqref{eq dmui SM} and \eqref{eq:dnui SM}, we take their ratio
\begin{equation}
  \frac{d \nu_i}{d \mu_i} = -2 k \phi \mu_i + \frac{\nu_i}{\mu_i}.
  \label{eq dnudmu}
\end{equation}
This is solved by
\begin{equation}
  \nu_i = -2 k \phi\mu_i^2 + C \mu_i.
  \label{eq nu(mu)}
\end{equation}
The constant $C$ is set by initial conditions. At the beginning of the winter, a fraction $q$ of the population is exposed to the disease and, a fraction $\mu_r$ of the population being susceptible, we have $\mu_i(0) = 1 - q\mu_r$. The initial density infected regions $\nu_i(0)$ is instead computed as follows. A given vertex will be an edge between an infected region and a non-infected one with probability $2q\mu_r(1-q\mu_r)$. Because each continuous non-infected region is associated with two such edges, we get $\nu_i(0) = q\mu_r(1-q\mu_r)$. The constant $C$ therefore reads
\begin{equation}
  C = q\mu_r + 2 k \phi(1-q\mu_r).
  \label{eq C}
\end{equation}
Plugging Eq.~\eqref{eq nu(mu)} back into Eq.~\eqref{eq dmui SM} we get
\begin{equation}
  \frac{d\mu_i}{2k \left(2k\phi\mu_i^2 - C\mu_i\right) \left( 1 - \frac{1-\mu_r}{\mu_i}\right)} = dt,
  \label{eq dmu bis}
\end{equation}
from which, after completing the square and integrating both sides, we get
\begin{equation}
  \frac{4}{\gamma^2} \int_{\mu_i(0)}^{\mu_i(t)} \frac{dz}{\left(\frac{2z - \alpha}{\gamma}\right)^2 - 1} = 4 k^2 \phi t,
  \label{eq int}
\end{equation}
where we defined the positive constants
\begin{equation}
  \alpha = \frac{C}{2k\phi} + (1-\mu_r);
  \qquad
  \gamma = \frac{C}{2k\phi} - (1-\mu_r).
\end{equation}
Solving the integral in Eq.~\eqref{eq int}, we get
\begin{equation}
  4 k^2 \phi t = \frac{1}{\gamma} \log \left(\frac{1+x(0)}{1-x(0)} \frac{1-x(t)}{1+x(t)}\right),
  \label{eq t vs x}
\end{equation}
where $x(t) = \frac{2\mu_i(t) - \alpha}{\gamma}$. Inverting Eq.~\eqref{eq t vs x} for $x(t)$ we obtain
\begin{equation}
  x(t) = \frac{x(0) - \tanh(2\gamma k\tilde{k} \phi t)}{1 - x(0) \tanh(2\gamma k\tilde{k} \phi t)}.
\end{equation}
From this, we finally obtain the fraction of infected people $p_i(t) = 1 - \mu_i(t)$ for a given fraction of immune people $p_r = 1 - \mu_r$ throughout a winter
\begin{equation}
  p_i(t) = (1-p_r) \frac
  {q(\psi + 1 - q) + \left[2\psi - q(\psi + 1 - q)\right] \tanh(\theta t)}
  {\psi + 1 - q + (\psi - 1 + q) \tanh(\theta t)},
  \label{eq pi(t)}
\end{equation}
in which
\begin{equation}
  \psi = \frac{q}{2k\phi},
  \qquad
  \theta = 2 k^2 \phi \mu_r (1 - q + \psi).
\end{equation}
After straightforward manipulations, Eq.~\eqref{eq pi(t)} is rearranged as
\begin{equation}
  f(p_r, t) = (1-p_r) \frac
  {\psi e^{2\theta t} + (1-q)(q-\psi)}
  {\psi e^{2\theta t} + 1 - q}.
  \label{eq pi2(t)}
\end{equation}
In the case $q = \phi$, Eq.~\eqref{eq pi2(t)} recovers Eq.~\eqref{eq:mf} in the main text.

\section{Details on the numerics}
\label{app:numerics}
Here we describe the various numerical routines used to calculate the data shown in the main text.

\subsection{Single trace}
To calculate the infected fraction as a function of time, as for example displayed in \cref{fig:3}a, we follow these steps:
\begin{enumerate}
  \item
  	Randomly generate a small-world graph~\cite{Watts1998}. This is done by taking a periodic chain of $N$ vertices, each connected to its $1^{\mathrm{st}},2^{\mathrm{nd}},\cdots,k^{\mathrm{th}}$ neighbors (resulting in a coordination number $2k$), and moving each edge end to a random location with a probability $\phi$.
  	For the results shown in the main text, we take $k=2$ everywhere and vary $\phi$ and $N$ according to the description in the figures' captions.
  \item
  	To initialize the dynamics, each of the $N$ vertices is infected with a probability $q$ (Autumn in \cref{fig:1}c).
  	The parameter $q$ is in principle independent, but, since the autumn infections can for example be thought of as arising from long-range connections to other parts of the world, it is meaningful to consider $q \sim \phi$. For concreteness, in the main text we considered $q=\phi$.
  	We note that this is not essential for the phase transition to occur, and we have extensively checked other choices as well (\emph{e.g.}, fixed $q$, varying $\phi$ or \emph{vice versa}). 
  \item 
  	The infection dynamics is run for $T_w$ time steps, which yields a certain final state. The fraction of infected vertices at each point in time is denoted by $p_i(t)$, and this is what we show for instance in \cref{fig:3}a,b.
  \item
  	In ``summer'', all the vertices that were in the immune state are put back in the susceptible state, whereas the infected vertices are transferred to the immune state, such that they do not participate in the infection dynamics of the following year.
  	Points 2-4 are repeated for every year, generating time traces such as in \cref{fig:3}b.
\end{enumerate}

\subsection{Phase diagram}

To each year in a multi-time trace we can associate the total number of infected vertices in that year, which measures the size of the outbreak.
Corresponding sample data are shown in \cref{fig:S1}a.
Depending on system parameters, we observe that these values cluster either around a single or two values, which is easily appreciated by looking at histograms of the data (\cref{fig:S1}b).
We extract the dominant peaks from the histograms and by majority vote over many averages determine whether a given set of parameters lies in the trivial or DTC phase (\cref{fig:S1}c), which we assemble to form the phase diagram (\cref{fig:S1}d).

\subsection{Bifurcation diagram}

\begin{figure*}[tb]
  \centering
  \includegraphics[width=\linewidth]{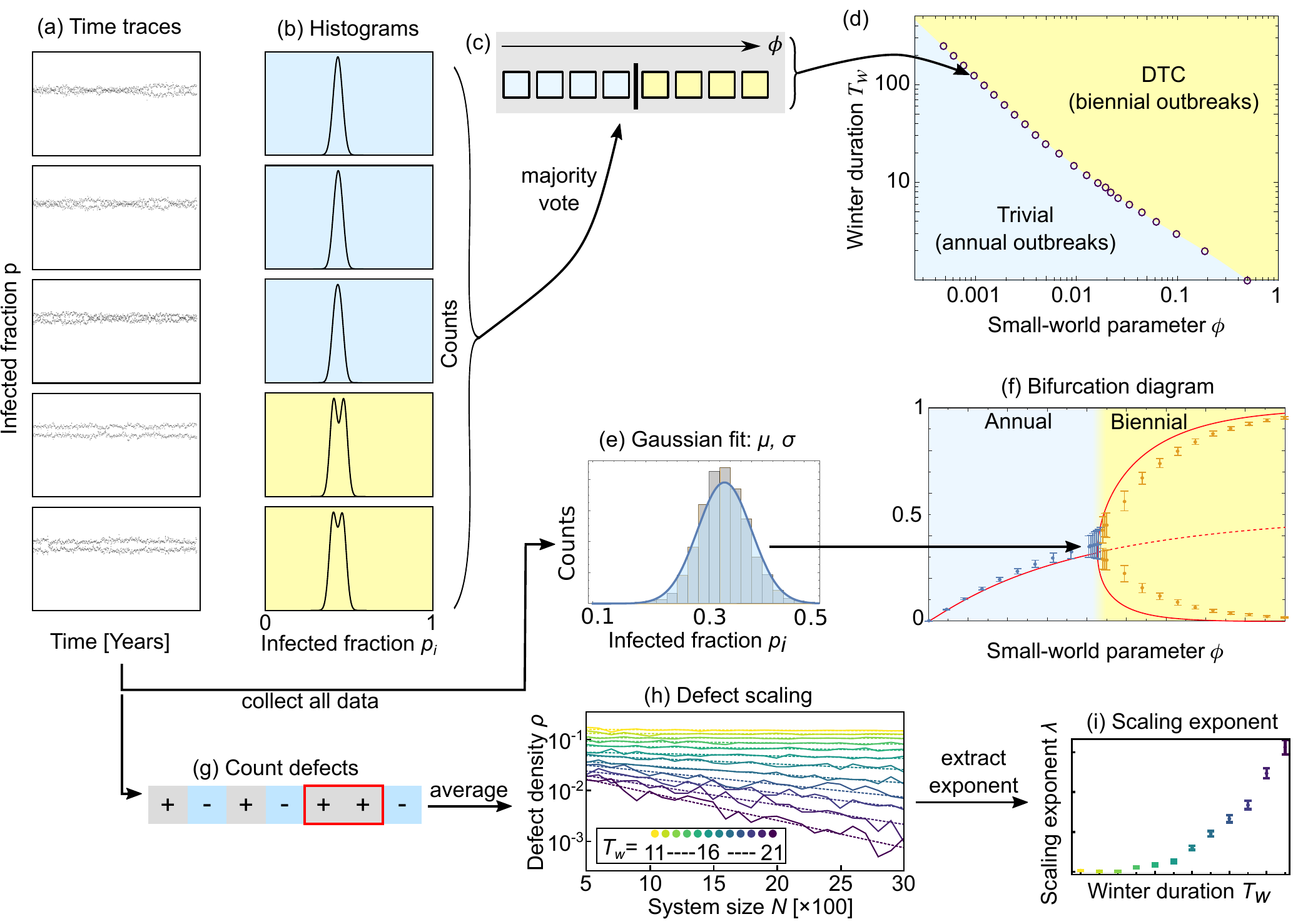}
  \caption{
  	\textbf{Data analysis.}
  	\textbf{(a)} For a given set of parameters $N,\phi,T_w,k$, we generate a large sample of random graphs and simulate long time traces (10,000 years).
  	From each time trace, we extract the fraction of infected vertices at the end of each year.
  	\textbf{(b)} These data are used to form histograms, which reveal either one (blue) or two (yellow) peaks, corresponding to annual and biennial epidemics, respectively.
  	Because of finite-size effects, close to the phase transition, fluctuations of the graphs mean that individual time traces can lie either in the trivial phase (blue) or in the DTC phase (yellow), which generates some uncertainty.
  	\textbf{(c)} By majority vote we extract from the collection histograms whether a set of parameters corresponds to the trivial or the DTC phase.
  	\textbf{(d)} From the classification (c) we estimate where the phase transition lies. The estimate is represented by a point in the phase diagram.
  	\textbf{(e)} For the best estimate of mean and standard deviation associated to one set of parameters, we collect all associated data and fit one or two Gaussians, depending on the phase it lies in.
  	\textbf{(f)} The computed mean and variance are shown as points with errors bars in the bifurcation diagram.
  	\textbf{(g)} The stability of the time crystal is assessed by counting defects.
  	\textbf{(h)} In the DTC phase the defect density decreases exponentially with system size, whereas in the trivial phase no change occurs with system size.
  	\textbf{(i)} We show the exponents obtained from fitting exponentials as a function of winter duration $T_w$.
  }
  \label{fig:S1}
\end{figure*}

To obtain a bifurcation diagram, we again take the data of a large number of time traces, but now fit either a single or two Gaussians to the total distribution of all infected fractions (\cref{fig:S1}e).
The obtained mean and variance are represented by a point and error bars in the bifurcation diagram (\cref{fig:S1}f).

\subsection{Defect density scaling}
We again take all time traces corresponding to a given set of parameters $\{N,\phi,T_w\}$ and now calculate the differences in the total infected fraction of adjacent years. 
In a DTC, the differences should alternate between positive and negative differences. We count the number of occurrences of differences of the same sign next to each other (\cref{fig:S1}g) and divide by the total number of years to obtain the defect density. 
This is calculated for $N=\{5,6,\cdots,30\}\times 10^2$ which reveals how the defect density changes with system size (\cref{fig:S1}h).
We observe clear exponential behavior, to which we fit exponentials.
The fitted exponents and errors are shown in a second diagram as function of $\phi$ for a specific value of $T_w$ (\cref{fig:S1}i). 

\clearpage

\bibliography{library,other_references}{}

\begin{thebibliography}{43}%
\makeatletter
\providecommand \@ifxundefined [1]{%
 \@ifx{#1\undefined}
}%
\providecommand \@ifnum [1]{%
 \ifnum #1\expandafter \@firstoftwo
 \else \expandafter \@secondoftwo
 \fi
}%
\providecommand \@ifx [1]{%
 \ifx #1\expandafter \@firstoftwo
 \else \expandafter \@secondoftwo
 \fi
}%
\providecommand \natexlab [1]{#1}%
\providecommand \enquote  [1]{``#1''}%
\providecommand \bibnamefont  [1]{#1}%
\providecommand \bibfnamefont [1]{#1}%
\providecommand \citenamefont [1]{#1}%
\providecommand \href@noop [0]{\@secondoftwo}%
\providecommand \href [0]{\begingroup \@sanitize@url \@href}%
\providecommand \@href[1]{\@@startlink{#1}\@@href}%
\providecommand \@@href[1]{\endgroup#1\@@endlink}%
\providecommand \@sanitize@url [0]{\catcode `\\12\catcode `\$12\catcode
  `\&12\catcode `\#12\catcode `\^12\catcode `\_12\catcode `\%12\relax}%
\providecommand \@@startlink[1]{}%
\providecommand \@@endlink[0]{}%
\providecommand \url  [0]{\begingroup\@sanitize@url \@url }%
\providecommand \@url [1]{\endgroup\@href {#1}{\urlprefix }}%
\providecommand \urlprefix  [0]{URL }%
\providecommand \Eprint [0]{\href }%
\providecommand \doibase [0]{http://dx.doi.org/}%
\providecommand \selectlanguage [0]{\@gobble}%
\providecommand \bibinfo  [0]{\@secondoftwo}%
\providecommand \bibfield  [0]{\@secondoftwo}%
\providecommand \translation [1]{[#1]}%
\providecommand \BibitemOpen [0]{}%
\providecommand \bibitemStop [0]{}%
\providecommand \bibitemNoStop [0]{.\EOS\space}%
\providecommand \EOS [0]{\spacefactor3000\relax}%
\providecommand \BibitemShut  [1]{\csname bibitem#1\endcsname}%
\let\auto@bib@innerbib\@empty
\bibitem [{\citenamefont {Bernoulli}(1766)}]{Bernoulli1760}%
  \BibitemOpen
  \bibfield  {author} {\bibinfo {author} {\bibfnamefont {Daniel}\ \bibnamefont
  {Bernoulli}},\ }\bibfield  {title} {\enquote {\bibinfo {title} {{Essai d'une
  nouvelle analyse de la mortalite causee par la petite verole}},}\ }\href@noop
  {} {\bibfield  {journal} {\bibinfo  {journal} {Mem Math Phy Acad Roy Sci
  Paris}\ } (\bibinfo {year} {1766})}\BibitemShut {NoStop}%
\bibitem [{\citenamefont {Anderson}\ and\ \citenamefont
  {May}(1992)}]{Anderson1992}%
  \BibitemOpen
  \bibfield  {author} {\bibinfo {author} {\bibfnamefont {Roy~M.}\ \bibnamefont
  {Anderson}}\ and\ \bibinfo {author} {\bibfnamefont {Robert~M.}\ \bibnamefont
  {May}},\ }\href
  {https://global.oup.com/academic/product/infectious-diseases-of-humans-9780198540403?cc=de&lang=en&}
  {\emph {\bibinfo {title} {{Infectious Diseases of Humans}}}}\ (\bibinfo
  {publisher} {Oxford University Press},\ \bibinfo {address} {Oxford},\
  \bibinfo {year} {1992})\ p.\ \bibinfo {pages} {766}\BibitemShut {NoStop}%
\bibitem [{\citenamefont {Keeling}\ and\ \citenamefont
  {Rohani}(2011)}]{KeelingAndRohani2011}%
  \BibitemOpen
  \bibfield  {author} {\bibinfo {author} {\bibfnamefont {Matt~J}\ \bibnamefont
  {Keeling}}\ and\ \bibinfo {author} {\bibfnamefont {Pejman}\ \bibnamefont
  {Rohani}},\ }\href {\doibase 10.2307/j.ctvcm4gk0} {\emph {\bibinfo {title}
  {{Modeling Infectious Diseases in Humans and Animals}}}}\ (\bibinfo
  {publisher} {Princeton University Press},\ \bibinfo {year}
  {2011})\BibitemShut {NoStop}%
\bibitem [{\citenamefont {Hamer}(1906)}]{Hamer1906}%
  \BibitemOpen
  \bibfield  {author} {\bibinfo {author} {\bibfnamefont {W.~H.}\ \bibnamefont
  {Hamer}},\ }\bibfield  {title} {\enquote {\bibinfo {title} {{On epidemic
  disease in England---The evidence of variability and persistency of type}},}\
  }\href {\doibase 10.1016/S0140-6736(01)80340-8} {\bibfield  {journal}
  {\bibinfo  {journal} {The Lancet}\ }\textbf {\bibinfo {volume} {167}},\
  \bibinfo {pages} {733--739} (\bibinfo {year} {1906})}\BibitemShut {NoStop}%
\bibitem [{\citenamefont {Kermack}\ and\ \citenamefont
  {McKendrick}(1927)}]{Kermack1927}%
  \BibitemOpen
  \bibfield  {author} {\bibinfo {author} {\bibfnamefont {W.~O.;}\ \bibnamefont
  {Kermack}}\ and\ \bibinfo {author} {\bibfnamefont {A.~G.}\ \bibnamefont
  {McKendrick}},\ }\bibfield  {title} {\enquote {\bibinfo {title} {{A
  contribution to the mathematical theory of epidemics}},}\ }\href {\doibase
  10.1098/rspa.1927.0118} {\bibfield  {journal} {\bibinfo  {journal}
  {Proceedings of the Royal Society of London. Series A, Containing Papers of a
  Mathematical and Physical Character}\ }\textbf {\bibinfo {volume} {115}},\
  \bibinfo {pages} {700--721} (\bibinfo {year} {1927})}\BibitemShut {NoStop}%
\bibitem [{\citenamefont {Soper}(1929)}]{Soper1929}%
  \BibitemOpen
  \bibfield  {author} {\bibinfo {author} {\bibfnamefont {H.~E.}\ \bibnamefont
  {Soper}},\ }\bibfield  {title} {\enquote {\bibinfo {title} {{The
  Interpretation of Periodicity in Disease Prevalence}},}\ }\href {\doibase
  10.2307/2341437} {\bibfield  {journal} {\bibinfo  {journal} {Journal of the
  Royal Statistical Society}\ }\textbf {\bibinfo {volume} {92}},\ \bibinfo
  {pages} {34} (\bibinfo {year} {1929})}\BibitemShut {NoStop}%
\bibitem [{\citenamefont {London}\ and\ \citenamefont
  {Yorke}(1973)}]{London1973}%
  \BibitemOpen
  \bibfield  {author} {\bibinfo {author} {\bibfnamefont {Wayne~P.}\
  \bibnamefont {London}}\ and\ \bibinfo {author} {\bibfnamefont {James~A.}\
  \bibnamefont {Yorke}},\ }\bibfield  {title} {\enquote {\bibinfo {title}
  {{Recurrent outbreaks of measles, chickenpox, and mumps}},}\ }\href {\doibase
  10.1093/oxfordjournals.aje.a121575} {\bibfield  {journal} {\bibinfo
  {journal} {American Journal of Epidemiology}\ }\textbf {\bibinfo {volume}
  {98}},\ \bibinfo {pages} {453--468} (\bibinfo {year} {1973})}\BibitemShut
  {NoStop}%
\bibitem [{\citenamefont {Kissler}\ \emph {et~al.}(2020)\citenamefont
  {Kissler}, \citenamefont {Tedijanto}, \citenamefont {Goldstein},
  \citenamefont {Grad},\ and\ \citenamefont {Lipsitch}}]{Kissler2020}%
  \BibitemOpen
  \bibfield  {author} {\bibinfo {author} {\bibfnamefont {Stephen~M.}\
  \bibnamefont {Kissler}}, \bibinfo {author} {\bibfnamefont {Christine}\
  \bibnamefont {Tedijanto}}, \bibinfo {author} {\bibfnamefont {Edward}\
  \bibnamefont {Goldstein}}, \bibinfo {author} {\bibfnamefont {Yonatan~H.}\
  \bibnamefont {Grad}}, \ and\ \bibinfo {author} {\bibfnamefont {Marc}\
  \bibnamefont {Lipsitch}},\ }\bibfield  {title} {\enquote {\bibinfo {title}
  {{Projecting the transmission dynamics of SARS-CoV-2 through the postpandemic
  period}},}\ }\href {\doibase 10.1126/science.abb5793} {\bibfield  {journal}
  {\bibinfo  {journal} {Science}\ }\textbf {\bibinfo {volume} {368}},\ \bibinfo
  {pages} {860--868} (\bibinfo {year} {2020})}\BibitemShut {NoStop}%
\bibitem [{\citenamefont {Marx}\ \emph
  {et~al.}(1997{\natexlab{a}})\citenamefont {Marx}, \citenamefont
  {T{\"{o}}r{\"{o}}k}, \citenamefont {Holman}, \citenamefont {Clarke},\ and\
  \citenamefont {Anderson}}]{Marx1997}%
  \BibitemOpen
  \bibfield  {author} {\bibinfo {author} {\bibfnamefont {Arthur}\ \bibnamefont
  {Marx}}, \bibinfo {author} {\bibfnamefont {Thomas~J.}\ \bibnamefont
  {T{\"{o}}r{\"{o}}k}}, \bibinfo {author} {\bibfnamefont {Robert~C.}\
  \bibnamefont {Holman}}, \bibinfo {author} {\bibfnamefont {Matthew~J.}\
  \bibnamefont {Clarke}}, \ and\ \bibinfo {author} {\bibfnamefont {Larry~J.}\
  \bibnamefont {Anderson}},\ }\bibfield  {title} {\enquote {\bibinfo {title}
  {{Pediatric Hospitalizations for Croup (Laryngotracheobronchitis): Biennial
  Increases Associated with Human Parainfluenza Virus 1 Epidemics}},}\ }\href
  {\doibase 10.1086/514137} {\bibfield  {journal} {\bibinfo  {journal} {The
  Journal of Infectious Diseases}\ }\textbf {\bibinfo {volume} {176}},\
  \bibinfo {pages} {1423--1427} (\bibinfo {year}
  {1997}{\natexlab{a}})}\BibitemShut {NoStop}%
\bibitem [{\citenamefont {Marx}\ \emph
  {et~al.}(1997{\natexlab{b}})\citenamefont {Marx}, \citenamefont
  {T{\"o}r{\"o}k}, \citenamefont {Holma}, \citenamefont {Clarke},\ and\
  \citenamefont {Anderson}}]{marx1997pediatric}%
  \BibitemOpen
  \bibfield  {author} {\bibinfo {author} {\bibfnamefont {Arthur}\ \bibnamefont
  {Marx}}, \bibinfo {author} {\bibfnamefont {Thomas~J}\ \bibnamefont
  {T{\"o}r{\"o}k}}, \bibinfo {author} {\bibfnamefont {Robert~C}\ \bibnamefont
  {Holma}}, \bibinfo {author} {\bibfnamefont {Matthew~J}\ \bibnamefont
  {Clarke}}, \ and\ \bibinfo {author} {\bibfnamefont {Larry~J}\ \bibnamefont
  {Anderson}},\ }\bibfield  {title} {\enquote {\bibinfo {title} {Pediatric
  hospitalizations for croup (laryngotracheobronchitis): biennial increases
  associated with human parainfluenza virus 1 epidemics},}\ }\href@noop {}
  {\bibfield  {journal} {\bibinfo  {journal} {Journal of Infectious Diseases}\
  }\textbf {\bibinfo {volume} {176}},\ \bibinfo {pages} {1423--1427} (\bibinfo
  {year} {1997}{\natexlab{b}})}\BibitemShut {NoStop}%
\bibitem [{\citenamefont {{Van Mieghem}}(2014)}]{VanMieghem2014}%
  \BibitemOpen
  \bibfield  {author} {\bibinfo {author} {\bibfnamefont {Piet}\ \bibnamefont
  {{Van Mieghem}}},\ }\href {\doibase 10.1017/CBO9781107415874} {\emph
  {\bibinfo {title} {{Performance Analysis of Complex Networks and Systems}}}}\
  (\bibinfo  {publisher} {Cambridge University Press},\ \bibinfo {address}
  {Cambridge},\ \bibinfo {year} {2014})\BibitemShut {NoStop}%
\bibitem [{\citenamefont {Pastor-Satorras}\ \emph {et~al.}(2015)\citenamefont
  {Pastor-Satorras}, \citenamefont {Castellano}, \citenamefont {{Van
  Mieghem}},\ and\ \citenamefont {Vespignani}}]{Pastor-Satorras2015}%
  \BibitemOpen
  \bibfield  {author} {\bibinfo {author} {\bibfnamefont {Romualdo}\
  \bibnamefont {Pastor-Satorras}}, \bibinfo {author} {\bibfnamefont {Claudio}\
  \bibnamefont {Castellano}}, \bibinfo {author} {\bibfnamefont {Piet}\
  \bibnamefont {{Van Mieghem}}}, \ and\ \bibinfo {author} {\bibfnamefont
  {Alessandro}\ \bibnamefont {Vespignani}},\ }\bibfield  {title} {\enquote
  {\bibinfo {title} {{Epidemic processes in complex networks}},}\ }\href
  {\doibase 10.1103/RevModPhys.87.925} {\bibfield  {journal} {\bibinfo
  {journal} {Reviews of Modern Physics}\ }\textbf {\bibinfo {volume} {87}},\
  \bibinfo {pages} {925--979} (\bibinfo {year} {2015})}\BibitemShut {NoStop}%
\bibitem [{\citenamefont {Vespignani}(2012)}]{Vespignani2012}%
  \BibitemOpen
  \bibfield  {author} {\bibinfo {author} {\bibfnamefont {Alessandro}\
  \bibnamefont {Vespignani}},\ }\bibfield  {title} {\enquote {\bibinfo {title}
  {{Modelling dynamical processes in complex socio-technical systems}},}\
  }\href {\doibase 10.1038/nphys2160} {\bibfield  {journal} {\bibinfo
  {journal} {Nature Physics}\ }\textbf {\bibinfo {volume} {8}},\ \bibinfo
  {pages} {32--39} (\bibinfo {year} {2012})}\BibitemShut {NoStop}%
\bibitem [{\citenamefont {Cirillo}\ and\ \citenamefont
  {Taleb}(2020)}]{Cirillo2020}%
  \BibitemOpen
  \bibfield  {author} {\bibinfo {author} {\bibfnamefont {Pasquale}\
  \bibnamefont {Cirillo}}\ and\ \bibinfo {author} {\bibfnamefont
  {Nassim~Nicholas}\ \bibnamefont {Taleb}},\ }\bibfield  {title} {\enquote
  {\bibinfo {title} {{Tail risk of contagious diseases}},}\ }\href {\doibase
  10.1038/s41567-020-0921-x} {\bibfield  {journal} {\bibinfo  {journal} {Nature
  Physics}\ }\textbf {\bibinfo {volume} {16}},\ \bibinfo {pages} {606--613}
  (\bibinfo {year} {2020})}\BibitemShut {NoStop}%
\bibitem [{\citenamefont {Barab{\'{a}}si}\ and\ \citenamefont
  {Albert}(1999)}]{Barabasi1999}%
  \BibitemOpen
  \bibfield  {author} {\bibinfo {author} {\bibfnamefont
  {Albert-L{\'{a}}szl{\'{o}}}\ \bibnamefont {Barab{\'{a}}si}}\ and\ \bibinfo
  {author} {\bibfnamefont {R{\'{e}}ka}\ \bibnamefont {Albert}},\ }\bibfield
  {title} {\enquote {\bibinfo {title} {{Emergence of Scaling in Random
  Networks}},}\ }\href {\doibase 10.1126/science.286.5439.509} {\bibfield
  {journal} {\bibinfo  {journal} {Science}\ }\textbf {\bibinfo {volume}
  {286}},\ \bibinfo {pages} {509--512} (\bibinfo {year} {1999})}\BibitemShut
  {NoStop}%
\bibitem [{\citenamefont {Pastor-Satorras}\ and\ \citenamefont
  {Vespignani}(2001)}]{Pastor-Satorras2001}%
  \BibitemOpen
  \bibfield  {author} {\bibinfo {author} {\bibfnamefont {Romualdo}\
  \bibnamefont {Pastor-Satorras}}\ and\ \bibinfo {author} {\bibfnamefont
  {Alessandro}\ \bibnamefont {Vespignani}},\ }\bibfield  {title} {\enquote
  {\bibinfo {title} {{Epidemic Spreading in Scale-Free Networks}},}\ }\href
  {\doibase 10.1103/PhysRevLett.86.3200} {\bibfield  {journal} {\bibinfo
  {journal} {Physical Review Letters}\ }\textbf {\bibinfo {volume} {86}},\
  \bibinfo {pages} {3200--3203} (\bibinfo {year} {2001})}\BibitemShut {NoStop}%
\bibitem [{\citenamefont {May}\ and\ \citenamefont {Lloyd}(2001)}]{May2001}%
  \BibitemOpen
  \bibfield  {author} {\bibinfo {author} {\bibfnamefont {Robert~M.}\
  \bibnamefont {May}}\ and\ \bibinfo {author} {\bibfnamefont {Alun~L.}\
  \bibnamefont {Lloyd}},\ }\bibfield  {title} {\enquote {\bibinfo {title}
  {{Infection dynamics on scale-free networks}},}\ }\href {\doibase
  10.1103/PhysRevE.64.066112} {\bibfield  {journal} {\bibinfo  {journal}
  {Physical Review E}\ }\textbf {\bibinfo {volume} {64}},\ \bibinfo {pages}
  {066112} (\bibinfo {year} {2001})}\BibitemShut {NoStop}%
\bibitem [{\citenamefont {Pastor-Satorras}\ and\ \citenamefont
  {Vespignani}(2002)}]{Pastor-Satorras2002}%
  \BibitemOpen
  \bibfield  {author} {\bibinfo {author} {\bibfnamefont {Romualdo}\
  \bibnamefont {Pastor-Satorras}}\ and\ \bibinfo {author} {\bibfnamefont
  {Alessandro}\ \bibnamefont {Vespignani}},\ }\bibfield  {title} {\enquote
  {\bibinfo {title} {{Epidemic dynamics in finite size scale-free networks}},}\
  }\href {\doibase 10.1103/PhysRevE.65.035108} {\bibfield  {journal} {\bibinfo
  {journal} {Physical Review E}\ }\textbf {\bibinfo {volume} {65}},\ \bibinfo
  {pages} {035108} (\bibinfo {year} {2002})}\BibitemShut {NoStop}%
\bibitem [{\citenamefont {Goldenberg}\ \emph {et~al.}(2005)\citenamefont
  {Goldenberg}, \citenamefont {Shavitt}, \citenamefont {Shir},\ and\
  \citenamefont {Solomon}}]{Goldenberg2005}%
  \BibitemOpen
  \bibfield  {author} {\bibinfo {author} {\bibfnamefont {Jacob}\ \bibnamefont
  {Goldenberg}}, \bibinfo {author} {\bibfnamefont {Yuval}\ \bibnamefont
  {Shavitt}}, \bibinfo {author} {\bibfnamefont {Eran}\ \bibnamefont {Shir}}, \
  and\ \bibinfo {author} {\bibfnamefont {Sorin}\ \bibnamefont {Solomon}},\
  }\bibfield  {title} {\enquote {\bibinfo {title} {{Distributive immunization
  of networks against viruses using the ‘honey-pot' architecture}},}\ }\href
  {\doibase 10.1038/nphys177} {\bibfield  {journal} {\bibinfo  {journal}
  {Nature Physics}\ }\textbf {\bibinfo {volume} {1}},\ \bibinfo {pages}
  {184--188} (\bibinfo {year} {2005})}\BibitemShut {NoStop}%
\bibitem [{\citenamefont {Moore}\ and\ \citenamefont
  {Newman}(2000)}]{Moore2000}%
  \BibitemOpen
  \bibfield  {author} {\bibinfo {author} {\bibfnamefont {Cristopher}\
  \bibnamefont {Moore}}\ and\ \bibinfo {author} {\bibfnamefont {M.~E.~J.}\
  \bibnamefont {Newman}},\ }\bibfield  {title} {\enquote {\bibinfo {title}
  {{Epidemics and percolation in small-world networks}},}\ }\href {\doibase
  10.1103/PhysRevE.61.5678} {\bibfield  {journal} {\bibinfo  {journal}
  {Physical Review E}\ }\textbf {\bibinfo {volume} {61}},\ \bibinfo {pages}
  {5678--5682} (\bibinfo {year} {2000})}\BibitemShut {NoStop}%
\bibitem [{\citenamefont {Travers}\ and\ \citenamefont
  {Milgram}(1969)}]{Milgram1967}%
  \BibitemOpen
  \bibfield  {author} {\bibinfo {author} {\bibfnamefont {Jeffrey}\ \bibnamefont
  {Travers}}\ and\ \bibinfo {author} {\bibfnamefont {Stanley}\ \bibnamefont
  {Milgram}},\ }\bibfield  {title} {\enquote {\bibinfo {title} {{An
  Experimental Study of the Small World Problem}},}\ }\href {\doibase
  10.2307/2786545} {\bibfield  {journal} {\bibinfo  {journal} {Sociometry}\
  }\textbf {\bibinfo {volume} {32}},\ \bibinfo {pages} {425} (\bibinfo {year}
  {1969})}\BibitemShut {NoStop}%
\bibitem [{\citenamefont {Watts}\ and\ \citenamefont
  {Strogatz}(1998)}]{Watts1998}%
  \BibitemOpen
  \bibfield  {author} {\bibinfo {author} {\bibfnamefont {Duncan~J}\
  \bibnamefont {Watts}}\ and\ \bibinfo {author} {\bibfnamefont {Steven~H}\
  \bibnamefont {Strogatz}},\ }\bibfield  {title} {\enquote {\bibinfo {title}
  {{Collective dynamics of ‘small-world' networks}},}\ }\href {\doibase
  10.1038/30918} {\bibfield  {journal} {\bibinfo  {journal} {Nature}\ }\textbf
  {\bibinfo {volume} {393}},\ \bibinfo {pages} {440--442} (\bibinfo {year}
  {1998})}\BibitemShut {NoStop}%
\bibitem [{\citenamefont {Newman}\ \emph {et~al.}(2000)\citenamefont {Newman},
  \citenamefont {Moore},\ and\ \citenamefont {Watts}}]{Newman2000}%
  \BibitemOpen
  \bibfield  {author} {\bibinfo {author} {\bibfnamefont {M.~E.~J.}\
  \bibnamefont {Newman}}, \bibinfo {author} {\bibfnamefont {C.}~\bibnamefont
  {Moore}}, \ and\ \bibinfo {author} {\bibfnamefont {D.~J.}\ \bibnamefont
  {Watts}},\ }\bibfield  {title} {\enquote {\bibinfo {title} {{Mean-Field
  Solution of the Small-World Network Model}},}\ }\href {\doibase
  10.1103/PhysRevLett.84.3201} {\bibfield  {journal} {\bibinfo  {journal}
  {Physical Review Letters}\ }\textbf {\bibinfo {volume} {84}},\ \bibinfo
  {pages} {3201--3204} (\bibinfo {year} {2000})}\BibitemShut {NoStop}%
\bibitem [{\citenamefont {Hastings}(2003)}]{Hastings2003}%
  \BibitemOpen
  \bibfield  {author} {\bibinfo {author} {\bibfnamefont {M.~B.}\ \bibnamefont
  {Hastings}},\ }\bibfield  {title} {\enquote {\bibinfo {title} {{Mean-Field
  and Anomalous Behavior on a Small-World Network}},}\ }\href {\doibase
  10.1103/PhysRevLett.91.098701} {\bibfield  {journal} {\bibinfo  {journal}
  {Physical Review Letters}\ }\textbf {\bibinfo {volume} {91}},\ \bibinfo
  {pages} {098701} (\bibinfo {year} {2003})}\BibitemShut {NoStop}%
\bibitem [{\citenamefont {Khemani}\ \emph {et~al.}(2016)\citenamefont
  {Khemani}, \citenamefont {Lazarides}, \citenamefont {Moessner},\ and\
  \citenamefont {Sondhi}}]{Khemani2016}%
  \BibitemOpen
  \bibfield  {author} {\bibinfo {author} {\bibfnamefont {Vedika}\ \bibnamefont
  {Khemani}}, \bibinfo {author} {\bibfnamefont {Achilleas}\ \bibnamefont
  {Lazarides}}, \bibinfo {author} {\bibfnamefont {Roderich}\ \bibnamefont
  {Moessner}}, \ and\ \bibinfo {author} {\bibfnamefont {S.~L.}\ \bibnamefont
  {Sondhi}},\ }\bibfield  {title} {\enquote {\bibinfo {title} {{Phase Structure
  of Driven Quantum Systems}},}\ }\href {\doibase
  10.1103/PhysRevLett.116.250401} {\bibfield  {journal} {\bibinfo  {journal}
  {Physical Review Letters}\ }\textbf {\bibinfo {volume} {116}},\ \bibinfo
  {pages} {250401} (\bibinfo {year} {2016})}\BibitemShut {NoStop}%
\bibitem [{\citenamefont {Else}\ \emph {et~al.}(2016)\citenamefont {Else},
  \citenamefont {Bauer},\ and\ \citenamefont {Nayak}}]{Else2016}%
  \BibitemOpen
  \bibfield  {author} {\bibinfo {author} {\bibfnamefont {Dominic~V.}\
  \bibnamefont {Else}}, \bibinfo {author} {\bibfnamefont {Bela}\ \bibnamefont
  {Bauer}}, \ and\ \bibinfo {author} {\bibfnamefont {Chetan}\ \bibnamefont
  {Nayak}},\ }\bibfield  {title} {\enquote {\bibinfo {title} {{Floquet Time
  Crystals}},}\ }\href {\doibase 10.1103/PhysRevLett.117.090402} {\bibfield
  {journal} {\bibinfo  {journal} {Physical Review Letters}\ }\textbf {\bibinfo
  {volume} {117}},\ \bibinfo {pages} {090402} (\bibinfo {year}
  {2016})}\BibitemShut {NoStop}%
\bibitem [{\citenamefont {Yao}\ \emph {et~al.}(2017)\citenamefont {Yao},
  \citenamefont {Potter}, \citenamefont {Potirniche},\ and\ \citenamefont
  {Vishwanath}}]{Yao2017}%
  \BibitemOpen
  \bibfield  {author} {\bibinfo {author} {\bibfnamefont {N.~Y.}\ \bibnamefont
  {Yao}}, \bibinfo {author} {\bibfnamefont {A.~C.}\ \bibnamefont {Potter}},
  \bibinfo {author} {\bibfnamefont {I.-D.}\ \bibnamefont {Potirniche}}, \ and\
  \bibinfo {author} {\bibfnamefont {A.}~\bibnamefont {Vishwanath}},\ }\bibfield
   {title} {\enquote {\bibinfo {title} {{Discrete Time Crystals: Rigidity,
  Criticality, and Realizations}},}\ }\href {\doibase
  10.1103/PhysRevLett.118.030401} {\bibfield  {journal} {\bibinfo  {journal}
  {Physical Review Letters}\ }\textbf {\bibinfo {volume} {118}},\ \bibinfo
  {pages} {030401} (\bibinfo {year} {2017})}\BibitemShut {NoStop}%
\bibitem [{\citenamefont {Moessner}\ and\ \citenamefont
  {Sondhi}(2017)}]{Moessner2017}%
  \BibitemOpen
  \bibfield  {author} {\bibinfo {author} {\bibfnamefont {R.}~\bibnamefont
  {Moessner}}\ and\ \bibinfo {author} {\bibfnamefont {S.~L.}\ \bibnamefont
  {Sondhi}},\ }\bibfield  {title} {\enquote {\bibinfo {title} {{Equilibration
  and order in quantum Floquet matter}},}\ }\href {\doibase 10.1038/nphys4106}
  {\bibfield  {journal} {\bibinfo  {journal} {Nature Physics}\ }\textbf
  {\bibinfo {volume} {13}},\ \bibinfo {pages} {424--428} (\bibinfo {year}
  {2017})}\BibitemShut {NoStop}%
\bibitem [{\citenamefont {Sacha}(2015)}]{Sacha2015}%
  \BibitemOpen
  \bibfield  {author} {\bibinfo {author} {\bibfnamefont {Krzysztof}\
  \bibnamefont {Sacha}},\ }\bibfield  {title} {\enquote {\bibinfo {title}
  {{Modeling spontaneous breaking of time-translation symmetry}},}\ }\href
  {\doibase 10.1103/PhysRevA.91.033617} {\bibfield  {journal} {\bibinfo
  {journal} {Physical Review A}\ }\textbf {\bibinfo {volume} {91}},\ \bibinfo
  {pages} {033617} (\bibinfo {year} {2015})}\BibitemShut {NoStop}%
\bibitem [{\citenamefont {Choi}\ \emph {et~al.}(2017)\citenamefont {Choi},
  \citenamefont {Choi}, \citenamefont {Landig}, \citenamefont {Kucsko},
  \citenamefont {Zhou}, \citenamefont {Isoya}, \citenamefont {Jelezko},
  \citenamefont {Onoda}, \citenamefont {Sumiya}, \citenamefont {Khemani},
  \citenamefont {von Keyserlingk}, \citenamefont {Yao}, \citenamefont
  {Demler},\ and\ \citenamefont {Lukin}}]{Choi2017}%
  \BibitemOpen
  \bibfield  {author} {\bibinfo {author} {\bibfnamefont {Soonwon}\ \bibnamefont
  {Choi}}, \bibinfo {author} {\bibfnamefont {Joonhee}\ \bibnamefont {Choi}},
  \bibinfo {author} {\bibfnamefont {Renate}\ \bibnamefont {Landig}}, \bibinfo
  {author} {\bibfnamefont {Georg}\ \bibnamefont {Kucsko}}, \bibinfo {author}
  {\bibfnamefont {Hengyun}\ \bibnamefont {Zhou}}, \bibinfo {author}
  {\bibfnamefont {Junichi}\ \bibnamefont {Isoya}}, \bibinfo {author}
  {\bibfnamefont {Fedor}\ \bibnamefont {Jelezko}}, \bibinfo {author}
  {\bibfnamefont {Shinobu}\ \bibnamefont {Onoda}}, \bibinfo {author}
  {\bibfnamefont {Hitoshi}\ \bibnamefont {Sumiya}}, \bibinfo {author}
  {\bibfnamefont {Vedika}\ \bibnamefont {Khemani}}, \bibinfo {author}
  {\bibfnamefont {Curt}\ \bibnamefont {von Keyserlingk}}, \bibinfo {author}
  {\bibfnamefont {Norman~Y.}\ \bibnamefont {Yao}}, \bibinfo {author}
  {\bibfnamefont {Eugene}\ \bibnamefont {Demler}}, \ and\ \bibinfo {author}
  {\bibfnamefont {Mikhail~D.}\ \bibnamefont {Lukin}},\ }\bibfield  {title}
  {\enquote {\bibinfo {title} {{Observation of discrete time-crystalline order
  in a disordered dipolar many-body system}},}\ }\href {\doibase
  10.1038/nature21426} {\bibfield  {journal} {\bibinfo  {journal} {Nature}\
  }\textbf {\bibinfo {volume} {543}},\ \bibinfo {pages} {221--225} (\bibinfo
  {year} {2017})}\BibitemShut {NoStop}%
\bibitem [{\citenamefont {Zhang}\ \emph {et~al.}(2017)\citenamefont {Zhang},
  \citenamefont {Hess}, \citenamefont {Kyprianidis}, \citenamefont {Becker},
  \citenamefont {Lee}, \citenamefont {Smith}, \citenamefont {Pagano},
  \citenamefont {Potirniche}, \citenamefont {Potter}, \citenamefont
  {Vishwanath}, \citenamefont {Yao},\ and\ \citenamefont {Monroe}}]{Zhang2017}%
  \BibitemOpen
  \bibfield  {author} {\bibinfo {author} {\bibfnamefont {J.}~\bibnamefont
  {Zhang}}, \bibinfo {author} {\bibfnamefont {P.~W.}\ \bibnamefont {Hess}},
  \bibinfo {author} {\bibfnamefont {A.}~\bibnamefont {Kyprianidis}}, \bibinfo
  {author} {\bibfnamefont {P.}~\bibnamefont {Becker}}, \bibinfo {author}
  {\bibfnamefont {A.}~\bibnamefont {Lee}}, \bibinfo {author} {\bibfnamefont
  {J.}~\bibnamefont {Smith}}, \bibinfo {author} {\bibfnamefont
  {G.}~\bibnamefont {Pagano}}, \bibinfo {author} {\bibfnamefont {I.-D.}\
  \bibnamefont {Potirniche}}, \bibinfo {author} {\bibfnamefont {A.~C.}\
  \bibnamefont {Potter}}, \bibinfo {author} {\bibfnamefont {A.}~\bibnamefont
  {Vishwanath}}, \bibinfo {author} {\bibfnamefont {N.~Y.}\ \bibnamefont {Yao}},
  \ and\ \bibinfo {author} {\bibfnamefont {C.}~\bibnamefont {Monroe}},\
  }\bibfield  {title} {\enquote {\bibinfo {title} {{Observation of a discrete
  time crystal}},}\ }\href {\doibase 10.1038/nature21413} {\bibfield  {journal}
  {\bibinfo  {journal} {Nature}\ }\textbf {\bibinfo {volume} {543}},\ \bibinfo
  {pages} {217--220} (\bibinfo {year} {2017})}\BibitemShut {NoStop}%
\bibitem [{\citenamefont {Gong}\ \emph {et~al.}(2018)\citenamefont {Gong},
  \citenamefont {Hamazaki},\ and\ \citenamefont {Ueda}}]{Gong2018a}%
  \BibitemOpen
  \bibfield  {author} {\bibinfo {author} {\bibfnamefont {Zongping}\
  \bibnamefont {Gong}}, \bibinfo {author} {\bibfnamefont {Ryusuke}\
  \bibnamefont {Hamazaki}}, \ and\ \bibinfo {author} {\bibfnamefont {Masahito}\
  \bibnamefont {Ueda}},\ }\bibfield  {title} {\enquote {\bibinfo {title}
  {{Discrete Time-Crystalline Order in Cavity and Circuit QED Systems}},}\
  }\href {\doibase 10.1103/PhysRevLett.120.040404} {\bibfield  {journal}
  {\bibinfo  {journal} {Physical Review Letters}\ }\textbf {\bibinfo {volume}
  {120}},\ \bibinfo {pages} {040404} (\bibinfo {year} {2018})}\BibitemShut
  {NoStop}%
\bibitem [{\citenamefont {Pizzi}\ \emph {et~al.}(2019)\citenamefont {Pizzi},
  \citenamefont {Knolle},\ and\ \citenamefont {Nunnenkamp}}]{Pizzi2019}%
  \BibitemOpen
  \bibfield  {author} {\bibinfo {author} {\bibfnamefont {Andrea}\ \bibnamefont
  {Pizzi}}, \bibinfo {author} {\bibfnamefont {Johannes}\ \bibnamefont
  {Knolle}}, \ and\ \bibinfo {author} {\bibfnamefont {Andreas}\ \bibnamefont
  {Nunnenkamp}},\ }\bibfield  {title} {\enquote {\bibinfo {title} {{Period-$n$
  Discrete Time Crystals and Quasicrystals with Ultracold Bosons}},}\ }\href
  {\doibase 10.1103/PhysRevLett.123.150601} {\bibfield  {journal} {\bibinfo
  {journal} {Physical Review Letters}\ }\textbf {\bibinfo {volume} {123}},\
  \bibinfo {pages} {150601} (\bibinfo {year} {2019})}\BibitemShut {NoStop}%
\bibitem [{\citenamefont {Machado}\ \emph {et~al.}(2020)\citenamefont
  {Machado}, \citenamefont {Else}, \citenamefont {Kahanamoku-Meyer},
  \citenamefont {Nayak},\ and\ \citenamefont {Yao}}]{Machado2020}%
  \BibitemOpen
  \bibfield  {author} {\bibinfo {author} {\bibfnamefont {Francisco}\
  \bibnamefont {Machado}}, \bibinfo {author} {\bibfnamefont {Dominic~V.}\
  \bibnamefont {Else}}, \bibinfo {author} {\bibfnamefont {Gregory~D.}\
  \bibnamefont {Kahanamoku-Meyer}}, \bibinfo {author} {\bibfnamefont {Chetan}\
  \bibnamefont {Nayak}}, \ and\ \bibinfo {author} {\bibfnamefont {Norman~Y.}\
  \bibnamefont {Yao}},\ }\bibfield  {title} {\enquote {\bibinfo {title}
  {{Long-Range Prethermal Phases of Nonequilibrium Matter}},}\ }\href {\doibase
  10.1103/PhysRevX.10.011043} {\bibfield  {journal} {\bibinfo  {journal}
  {Physical Review X}\ }\textbf {\bibinfo {volume} {10}},\ \bibinfo {pages}
  {011043} (\bibinfo {year} {2020})}\BibitemShut {NoStop}%
\bibitem [{\citenamefont {Yao}\ \emph {et~al.}(2020)\citenamefont {Yao},
  \citenamefont {Nayak}, \citenamefont {Balents},\ and\ \citenamefont
  {Zaletel}}]{Yao2020}%
  \BibitemOpen
  \bibfield  {author} {\bibinfo {author} {\bibfnamefont {Norman~Y.}\
  \bibnamefont {Yao}}, \bibinfo {author} {\bibfnamefont {Chetan}\ \bibnamefont
  {Nayak}}, \bibinfo {author} {\bibfnamefont {Leon}\ \bibnamefont {Balents}}, \
  and\ \bibinfo {author} {\bibfnamefont {Michael~P.}\ \bibnamefont {Zaletel}},\
  }\bibfield  {title} {\enquote {\bibinfo {title} {{Classical discrete time
  crystals}},}\ }\href {\doibase 10.1038/s41567-019-0782-3} {\bibfield
  {journal} {\bibinfo  {journal} {Nature Physics}\ }\textbf {\bibinfo {volume}
  {16}},\ \bibinfo {pages} {438--447} (\bibinfo {year} {2020})}\BibitemShut
  {NoStop}%
\bibitem [{\citenamefont {Pizzi}\ \emph {et~al.}(2020)\citenamefont {Pizzi},
  \citenamefont {Nunnenkamp},\ and\ \citenamefont {Knolle}}]{Pizzi2020}%
  \BibitemOpen
  \bibfield  {author} {\bibinfo {author} {\bibfnamefont {Andrea}\ \bibnamefont
  {Pizzi}}, \bibinfo {author} {\bibfnamefont {Andreas}\ \bibnamefont
  {Nunnenkamp}}, \ and\ \bibinfo {author} {\bibfnamefont {Johannes}\
  \bibnamefont {Knolle}},\ }\bibfield  {title} {\enquote {\bibinfo {title}
  {{Bistability and time crystals in long-ranged directed percolation}},}\
  }\href {http://arxiv.org/abs/2004.13034} {\bibfield  {journal} {\bibinfo
  {journal} {arXiv:2004.13034}\ } (\bibinfo {year} {2020})}\BibitemShut
  {NoStop}%
\bibitem [{\citenamefont {Gambetta}\ \emph
  {et~al.}(2019{\natexlab{a}})\citenamefont {Gambetta}, \citenamefont
  {Carollo}, \citenamefont {Lazarides}, \citenamefont {Lesanovsky},\ and\
  \citenamefont {Garrahan}}]{Gambetta2019}%
  \BibitemOpen
  \bibfield  {author} {\bibinfo {author} {\bibfnamefont {F.~M.}\ \bibnamefont
  {Gambetta}}, \bibinfo {author} {\bibfnamefont {F.}~\bibnamefont {Carollo}},
  \bibinfo {author} {\bibfnamefont {A.}~\bibnamefont {Lazarides}}, \bibinfo
  {author} {\bibfnamefont {I.}~\bibnamefont {Lesanovsky}}, \ and\ \bibinfo
  {author} {\bibfnamefont {J.~P.}\ \bibnamefont {Garrahan}},\ }\bibfield
  {title} {\enquote {\bibinfo {title} {{Classical stochastic discrete time
  crystals}},}\ }\href {\doibase 10.1103/PhysRevE.100.060105} {\bibfield
  {journal} {\bibinfo  {journal} {Physical Review E}\ }\textbf {\bibinfo
  {volume} {100}},\ \bibinfo {pages} {060105} (\bibinfo {year}
  {2019}{\natexlab{a}})}\BibitemShut {NoStop}%
\bibitem [{\citenamefont {Gros}(2013)}]{Gros2013}%
  \BibitemOpen
  \bibfield  {author} {\bibinfo {author} {\bibfnamefont {Claudius}\
  \bibnamefont {Gros}},\ }\href {\doibase 10.1007/978-3-642-36586-7} {\emph
  {\bibinfo {title} {Complex and Adaptive Dynamical Systems}}}\ (\bibinfo
  {publisher} {Springer Berlin Heidelberg},\ \bibinfo {address} {Berlin,
  Heidelberg},\ \bibinfo {year} {2013})\BibitemShut {NoStop}%
\bibitem [{\citenamefont {Gambetta}\ \emph
  {et~al.}(2019{\natexlab{b}})\citenamefont {Gambetta}, \citenamefont
  {Carollo}, \citenamefont {Lazarides}, \citenamefont {Lesanovsky},\ and\
  \citenamefont {Garrahan}}]{gambetta2019classical}%
  \BibitemOpen
  \bibfield  {author} {\bibinfo {author} {\bibfnamefont {FM}~\bibnamefont
  {Gambetta}}, \bibinfo {author} {\bibfnamefont {F}~\bibnamefont {Carollo}},
  \bibinfo {author} {\bibfnamefont {A}~\bibnamefont {Lazarides}}, \bibinfo
  {author} {\bibfnamefont {I}~\bibnamefont {Lesanovsky}}, \ and\ \bibinfo
  {author} {\bibfnamefont {JP}~\bibnamefont {Garrahan}},\ }\bibfield  {title}
  {\enquote {\bibinfo {title} {Classical stochastic discrete time crystals},}\
  }\href@noop {} {\bibfield  {journal} {\bibinfo  {journal} {Physical Review
  E}\ }\textbf {\bibinfo {volume} {100}},\ \bibinfo {pages} {060105} (\bibinfo
  {year} {2019}{\natexlab{b}})}\BibitemShut {NoStop}%
\bibitem [{\citenamefont {Bennett}\ \emph {et~al.}(1990)\citenamefont
  {Bennett}, \citenamefont {Grinstein}, \citenamefont {He}, \citenamefont
  {Jayaprakash},\ and\ \citenamefont {Mukamel}}]{Bennett1990}%
  \BibitemOpen
  \bibfield  {author} {\bibinfo {author} {\bibfnamefont {Charles~H.}\
  \bibnamefont {Bennett}}, \bibinfo {author} {\bibfnamefont {G.}~\bibnamefont
  {Grinstein}}, \bibinfo {author} {\bibfnamefont {Yu}~\bibnamefont {He}},
  \bibinfo {author} {\bibfnamefont {C.}~\bibnamefont {Jayaprakash}}, \ and\
  \bibinfo {author} {\bibfnamefont {David}\ \bibnamefont {Mukamel}},\
  }\bibfield  {title} {\enquote {\bibinfo {title} {{Stability of temporally
  periodic states of classical many-body systems}},}\ }\href {\doibase
  10.1103/PhysRevA.41.1932} {\bibfield  {journal} {\bibinfo  {journal}
  {Physical Review A}\ }\textbf {\bibinfo {volume} {41}},\ \bibinfo {pages}
  {1932--1935} (\bibinfo {year} {1990})}\BibitemShut {NoStop}%
\bibitem [{\citenamefont {Goles}\ \emph {et~al.}(2001)\citenamefont {Goles},
  \citenamefont {Schulz},\ and\ \citenamefont {Markus}}]{Goles2001}%
  \BibitemOpen
  \bibfield  {author} {\bibinfo {author} {\bibfnamefont {Eric}\ \bibnamefont
  {Goles}}, \bibinfo {author} {\bibfnamefont {Oliver}\ \bibnamefont {Schulz}},
  \ and\ \bibinfo {author} {\bibfnamefont {Mario}\ \bibnamefont {Markus}},\
  }\bibfield  {title} {\enquote {\bibinfo {title} {{Prime number selection of
  cycles in a predator-prey model}},}\ }\href {\doibase 10.1002/cplx.1040}
  {\bibfield  {journal} {\bibinfo  {journal} {Complexity}\ }\textbf {\bibinfo
  {volume} {6}},\ \bibinfo {pages} {33--38} (\bibinfo {year}
  {2001})}\BibitemShut {NoStop}%
\bibitem [{\citenamefont {Kuperman}\ and\ \citenamefont
  {Abramson}(2001)}]{Kuperman2001}%
  \BibitemOpen
  \bibfield  {author} {\bibinfo {author} {\bibfnamefont {Marcelo}\ \bibnamefont
  {Kuperman}}\ and\ \bibinfo {author} {\bibfnamefont {Guillermo}\ \bibnamefont
  {Abramson}},\ }\bibfield  {title} {\enquote {\bibinfo {title} {{Small World
  Effect in an Epidemiological Model}},}\ }\href {\doibase
  10.1103/PhysRevLett.86.2909} {\bibfield  {journal} {\bibinfo  {journal}
  {Physical Review Letters}\ }\textbf {\bibinfo {volume} {86}},\ \bibinfo
  {pages} {2909--2912} (\bibinfo {year} {2001})}\BibitemShut {NoStop}%
\bibitem [{\citenamefont {Watanabe}\ and\ \citenamefont
  {Oshikawa}(2015)}]{Watanabe2015}%
  \BibitemOpen
  \bibfield  {author} {\bibinfo {author} {\bibfnamefont {Haruki}\ \bibnamefont
  {Watanabe}}\ and\ \bibinfo {author} {\bibfnamefont {Masaki}\ \bibnamefont
  {Oshikawa}},\ }\bibfield  {title} {\enquote {\bibinfo {title} {{Absence of
  Quantum Time Crystals}},}\ }\href {\doibase 10.1103/PhysRevLett.114.251603}
  {\bibfield  {journal} {\bibinfo  {journal} {Physical Review Letters}\
  }\textbf {\bibinfo {volume} {114}},\ \bibinfo {pages} {251603} (\bibinfo
  {year} {2015})}\BibitemShut {NoStop}%
\end{thebibliography}%
\end{document}